\documentclass[12pt,a4paper]{article}
\usepackage{amssymb}
\usepackage{amsfonts}
\usepackage{amsmath}
\usepackage{amsthm}
\usepackage{bbm}
\usepackage{epsfig}
\usepackage{array}
\usepackage{booktabs}
\usepackage{graphicx}
\usepackage{theorem}
\usepackage{array}
\usepackage[numbers, sort]{natbib}
\usepackage{hyperref}
\usepackage{color}
\usepackage[toc, page]{appendix}
\allowdisplaybreaks[1]

\def\L{\mathcal L}
\def\e{\varepsilon}

\newcommand{\wt}{\widetilde}

\newcommand{\be}{\begin{equation}}
\newcommand{\ee}{\end{equation}}
\newcommand{\beq}{\begin{equation}}
\newcommand{\eeq}{\end{equation}}
\newcommand{\bea}{\begin{eqnarray}}
\newcommand{\eea}{\end{eqnarray}}

\makeatletter

\newcommand{\Rmnum}[1]{\expandafter\@slowromancap\romannumeral #1@}
\makeatother

\begin{document}

\def\a{\alpha}
\def\b{\beta}
\def\c{\chi}
\def\d{\delta}
\def\e{\epsilon}
\def\f{\phi}
\def\g{\gamma}
\def\h{\eta}
\def\i{\iota}
\def\j{\psi}
\def\k{\kappa}
\def\l{\lambda}
\def\m{\mu}
\def\n{\nu}
\def\o{\omega}
\def\p{\pi}
\def\q{\theta}
\def\r{\rho}
\def\s{\sigma}
\def\t{\tau}
\def\u{\upsilon}
\def\x{\xi}
\def\z{\zeta}
\def\D{\Delta}
\def\F{\Phi}
\def\G{\Gamma}
\def\J{\Psi}
\def\L{\Lambda}
\def\O{\Omega}
\def\P{\Pi}
\def\Q{\Theta}
\def\S{\Sigma}
\def\U{\Upsilon}
\def\X{\Xi}

\def\ve{\varepsilon}
\def\vf{\varphi}
\def\vr{\varrho}
\def\vs{\varsigma}
\def\vq{\vartheta}

\def\dg{\dagger}                                     
\def\ddg{\ddagger}                                   
\def\wt#1{\widetilde{#1}}                    
\def\mt{\widetilde{m}_1}
\def\mti{\widetilde{m}_i}
\def\rt{\widetilde{r}_1}
\def\mtt{\widetilde{m}_2}
\def\mttt{\widetilde{m}_3}
\def\rtt{\widetilde{r}_2}
\def\mb{\overline{m}}
\def\VEV#1{\left\langle #1\right\rangle}        
\def\be{\begin{equation}}
\def\ee{\end{equation}}
\def\ds{\displaystyle}
\def\ra{\rightarrow}

\def\bea{\begin{eqnarray}}
\def\eea{\end{eqnarray}}
\def\NO{\nonumber}
\def\Bar#1{\overline{#1}}


\def\pl#1#2#3{Phys.~Lett.~{\bf B {#1}} ({#2}) #3}
\def\np#1#2#3{Nucl.~Phys.~{\bf B {#1}} ({#2}) #3}
\def\prl#1#2#3{Phys.~Rev.~Lett.~{\bf #1} ({#2}) #3}
\def\pr#1#2#3{Phys.~Rev.~{\bf D {#1}} ({#2}) #3}
\def\zp#1#2#3{Z.~Phys.~{\bf C {#1}} ({#2}) #3}
\def\cqg#1#2#3{Class.~and Quantum Grav.~{\bf {#1}} ({#2}) #3}
\def\cmp#1#2#3{Commun.~Math.~Phys.~{\bf {#1}} ({#2}) #3}
\def\jmp#1#2#3{J.~Math.~Phys.~{\bf {#1}} ({#2}) #3}
\def\ap#1#2#3{Ann.~of Phys.~{\bf {#1}} ({#2}) #3}
\def\prep#1#2#3{Phys.~Rep.~{\bf {#1}C} ({#2}) #3}
\def\ptp#1#2#3{Progr.~Theor.~Phys.~{\bf {#1}} ({#2}) #3}
\def\ijmp#1#2#3{Int.~J.~Mod.~Phys.~{\bf A {#1}} ({#2}) #3}
\def\mpl#1#2#3{Mod.~Phys.~Lett.~{\bf A {#1}} ({#2}) #3}
\def\nc#1#2#3{Nuovo Cim.~{\bf {#1}} ({#2}) #3}
\def\ibid#1#2#3{{\it ibid.}~{\bf {#1}} ({#2}) #3}

\title{
\vspace*{15mm}
\bf A fuller flavour treatment of $N_2$-dominated leptogenesis}
\author{{\Large Stefan~Antusch$^a$, Pasquale~Di~Bari$^{b,c}$, David~A.~Jones$^b$, Steve~F.~King$^b$}
\\
$^a$
{\it\small Max-Planck-Institut f\"{u}r Physik}
{\it\small (Werner-Heisenberg-Institut)} \\
{\it\small F\"{o}hringer Ring 6, 80805 M\"{u}nchen, Germany}\\
$^b$
{\it\small School of Physics and Astronomy},{\it\small University of Southampton,}
{\it\small  Southampton, SO17 1BJ, U.K.}\\
$^c$
{\it\small Department of Physics and Astronomy},{\it\small University of Sussex,}
{\it\small  Brighton, BN1 9QH, U.K.}
}

\maketitle \thispagestyle{empty}

\vspace{-10mm}

\begin{abstract}
  We discuss $N_2$-dominated leptogenesis in the presence of flavour dependent effects that have hitherto been neglected, in particular the off-diagonal entries of the flavour coupling matrix that
  connects the total flavour asymmetries, distributed in different particle species,
  to the lepton and Higgs doublet asymmetries. We derive analytical formulae for the final asymmetry
  including the flavour coupling at the $N_2$-decay stage as well as at the stage of washout by the lightest
  right-handed neutrino $N_1$. Moreover, we point out that in general part of the electron and muon asymmetries
  (phantom terms), can completely escape the wash-out at the production
   and a total $B-L$ asymmetry can be generated by the lightest RH neutrino wash-out
   yielding so called phantom leptogenesis. However, the phantom terms are proportional
   to the initial $N_2$ abundance and  in particular they vanish for initial zero $N_2$-abundance.
  Taking any of these new effects into account can significantly modify the final asymmetry produced by the decays of
  the next-to-lightest RH neutrinos, opening up new interesting possibilities for $N_2$-dominated thermal leptogenesis.
  \end{abstract}

\newpage

\section{Introduction}

Leptogenesis \cite{fy} is based on a popular extension of the Standard Model,
where three right-handed (RH) neutrinos $N_{R i}$, with a Majorana mass term
$M$ and Yukawa couplings $h$, are added to the SM Lagrangian,
\begin{equation}\label{lagrangian}
\mathcal{L}= \mathcal{L}_{\rm SM} +i \overline{N}_{R i}\g_{\m}\partial^{\m} N_{Ri} -
h_{\a i} \overline{\ell}_{L\a} N_{R i} \tilde{\F} -
{1\over 2}\,M_i \overline{N}_{R i}^c \, N_{R i} +h.c.\quad (i=1,2,3,\quad \a=e,\m,\t) \, .
\end{equation}
 After spontaneous symmetry breaking, a Dirac mass term $m_D=v\,h$, is generated
by the vev $v=174$ GeV of the Higgs boson. In the see-saw limit, $M\gg m_D$,
the spectrum of neutrino mass eigenstates
splits in two sets: 3 very heavy neutrinos $N_1,N_2$ and $N_3$,
respectively with masses $M_1\leq M_2 \leq M_3$, almost coinciding with
the eigenvalues of $M$, and 3 light neutrinos with masses $m_1\leq m_2\leq m_3$,
the eigenvalues of the light neutrino mass matrix
given by the see-saw formula \cite{seesaw}
\be
m_{\nu}= - m_D\,{1\over M}\,m_D^T \, .
\ee
Neutrino oscillation experiments measure two neutrino mass-squared
differences. For normal schemes one has
$m^{\,2}_3-m_2^{\,2}=\Delta m^2_{\rm atm}$ and
$m^{\,2}_2-m_1^{\,2}=\Delta m^2_{\rm sol}$,
whereas for inverted schemes one has
$m^{\,2}_3-m_2^{\,2}=\Delta m^2_{\rm sol}$
and $m^{\,2}_2-m_1^{\,2}=\Delta m^2_{\rm atm}$.
For $m_1\gg m_{\rm atm} \equiv
\sqrt{\Delta m^2_{\rm atm}+\Delta m^2_{\rm sol}}=
(0.050\pm 0.001)\,{\rm eV}$ \cite{gonzalez}
the spectrum is quasi-degenerate, while for
$m_1\ll m_{\rm sol}\equiv \sqrt{\D m^2_{\rm sol}}
=(0.0088\pm 0.0001)\,{\rm eV}$ \cite{gonzalez}
it is fully hierarchical (normal or inverted).
The most stringent upper bound on the
absolute neutrino mass scale comes from
cosmological observations. Recently, quite a conservative
upper bound,
\be\label{bound}
m_1 < 0.2\,{\rm eV} \, \hspace{5mm} (95\%\, {\rm CL}) \, ,
\ee
has been obtained by the
WMAP collaboration combining CMB, baryon acoustic oscillations
and supernovae type Ia observations \cite{WMAP5}.

The $C\!P$ violating  decays of the RH neutrinos into lepton doublets
and Higgs bosons at temperatures $T\gtrsim 100\,{\rm GeV}$
generate a $B-L$ asymmetry one third of which, thanks to sphaleron processes,
ends up into a baryon asymmetry that can explain the observed
baryon asymmetry of the Universe.
This can be expressed in terms of the baryon-to-photon number ratio and
a precise measurement comes from the CMBR anisotropies observations of WMAP \cite{WMAP5},
\be\label{etaBobs}
\eta_B^{\rm CMB} = (6.2 \pm 0.15)\times 10^{-10} \, .
\ee
The predicted baryon-to-photon ratio $\eta_B$ is related to the final
value of the $(B-L)$ asymmetry $N^{\rm f}_{B-L}$ by the relation
\be\label{etaB}
\eta_B \simeq 0.96\times 10^{-2} N_{B-L}^{\rm f} \, ,
\ee
where we indicate with $N_X$ any particle number or asymmetry $X$ calculated in a portion
of co-moving volume containing one heavy neutrino in ultra-relativistic
thermal equilibrium, so that e.g. $N^{\rm eq}_{N_2}(T\gg M_2)=1$.

If one imposes that the RH neutrino mass spectrum is strongly hierarchical, then there are
two options for successful leptogenesis. A first one is given by the $N_1$-dominated
scenario, where the final asymmetry is dominated by the decays of the lightest RH neutrinos.
The main limitation of this scenario is that successful leptogenesis implies quite a
restrictive lower bound on the mass of the lightest RH neutrino. Imposing independence of
the final asymmetry of the initial RH neutrino abundance
and barring phase cancelations in the see-saw orthogonal matrix entries the lower
bound is given by \cite{di,cmb,flavorlep}
\be\label{lb}
M_1 \gtrsim 3 \times 10^9 \,{\rm GeV} \, .
\ee
This implies in turn a lower bound
$T_{\rm reh} \gtrsim 1.5 \times 10^{9}\,{\rm GeV}$ on the reheating temperature as well \cite{annals}
\footnote{For a discussion of flavour-dependent leptogenesis  in the supersymmetric seesaw scenario and the corresponding bounds on $M_1$ and $T_{\rm reh} $, see \cite{Antusch:2006cw,Antusch:2006gy}.}.
The lower bound Eq.~(\ref{lb}) is  typically not respected in models emerging from
 grand unified theories. It has therefore been
long thought that, within a minimal type I see-saw mechanism,
leptogenesis is not viable within these models \cite{branco}.

There is however a second option \cite{geometry}, namely the
$N_2$-dominated leptogenesis scenario, where the asymmetry is
dominantly produced from the decays of the next-to-lightest RH neutrinos. In this case
there is no lower bound on the lightest RH neutrino mass $M_1$. Instead this is replaced by a lower bound on the
next-to-lightest RH neutrino mass $M_2$
that still implies a lower bound on the reheating temperature.

There are two necessary conditions for a successful $N_2$-dominated leptogenesis scenario.
The first one is the presence of (at least) a third heavier  RH neutrino $N_3$ that  couples to $N_2$ in
order for the $C\!P$ asymmetries of $N_2$ not to be suppressed as $\propto (M_1/M_2)^2$.
The second necessary condition is to be able to circumvent the wash-out from the lightest RH neutrinos. There is a
particular choice of the see-saw parameters where these two conditions are maximally satisfied.
This corresponds to the limit where the lightest RH neutrino gets decoupled, as in
heavy sequential dominance, an example which we shall discuss later.
In this case the bound, $M_2\gtrsim 10^{10}\,{\rm GeV}$ when estimated without the inclusion of flavour effects,
is saturated. In this limit the wash-out from the lightest RH neutrinos
is totally absent and the $C\!P$ asymmetries of the $N_2$'s are maximal.

In order to have successful $N_2$-dominated leptogenesis for choices of the parameters
not necessarily close to this maximal case a crucial role is played
by lepton flavour effects \cite{vives}.
If $M_1\ll 10^9\,{\rm GeV}\ll M_2$, as we will assume, then
before the lightest RH neutrino wash-out is active,
the quantum states of the lepton doublets produced by $N_2$-decays
get fully incoherent in flavour space \cite{nardi1,flavoreffects1,zeno,decoherence1,decoherence2}. In this way the lightest RH neutrino wash-out acts separately on each flavour asymmetry and is then much less efficient \cite{vives}
\footnote{Notice that if $M_1\gg 10^{9}\,{\rm GeV}$ and $K_1\gg 1$ the wash-out from the lightest RH neutrino
can be still avoided thanks to heavy flavour effects \cite{bcst,nardi2}. However, throughout this paper we will always consider the case $M_1\ll 10^9\,{\rm GeV}$ which is more interesting with respect to leptogenesis in grand-unified theories.}.
It has then been shown recently that within this scenario
it is possible to have successful leptogenesis within models
emerging from $SO(10)$ grand-unified theories with interesting
potential predictions on the low energy parameters \cite{SO10}.
Therefore, the relevance of the $N_2$-dominated scenario has been gradually
increasing in the last years.

In this paper we discuss $N_2$-dominated leptogenesis in the presence of
flavour dependent effects that have hitherto been neglected, in particular the off-diagonal entries of the flavour coupling matrix that
connects the total flavour asymmetries, distributed in different particle species, to the lepton and Higgs doublet asymmetries.
We derive analytical formulae for the final asymmetry
including the flavour coupling at the $N_2$-decay stage as well as at the stage of washout by the lightest
RH neutrino $N_1$. We point out that in general part of the electron and muon asymmetries will
completely escape the wash-out at the production and a total $B-L$ asymmetry
can be generated by the lightest RH neutrino wash-out yielding so called phantom leptogenesis.
These contributions, that we call phantom terms, introduce however a
strong dependence on the initial conditions as we explain in detail.
Taking of all these new effects into account can enhance the final asymmetry produced by the decays of
the next-to-lightest RH neutrinos by orders of magnitude, opening up new interesting possibilities for $N_2$-dominated
thermal leptogenesis. We illustrate these effects for two models which describe realistic neutrino masses and mixing
based on sequential dominance.

The layout of the remainder of the paper is as follows.
In section 2 we discuss the production of the asymmetry from $N_2$-decays
and its subsequent thermal washout at similar temperatures.
In section 3 we discuss three flavour projection and the wash-out stage
at lower temperatures relevant to the lightest RH neutrino mass.
This is where the asymmetry which survives from $N_2$-decays and washout would
typically be expected to be washed out by the lightest RH neutrinos
in a flavour independent treatment, but which typically survives in a flavour-dependent
treatment. This conclusion is reinforced in the fuller flavour treatment
here making $N_2$ dominated leptogenesis even more relevant.
The fuller flavour effects of the $N_2$-dominated scenario are
encoded in a compact master formula presented at the end of this section and partly unpacked in
an Appendix. Section 4 applies this master formula to examples where the new effects arising from the flavour couplings
and phantom leptogenesis play a prominent role. We focus on examples where,
due to the considered effects, the flavour asymmetry produced dominantly in one
flavour can emerge as an asymmetry in a different flavour, a scenario we refer to
as the flavour swap scenario.

\section{Production of the asymmetry from $N_2$-decays and washout}

In the $N_2$-dominated scenario, with $M_2$ respecting the lower bound of $M_2\gtrsim 10^{10}\,{\rm GeV}$
and $M_1\ll 10^9\,{\rm GeV}$, one has to distinguish two stages in the calculation
of the asymmetry.  In a first {\em production stage}, at $T\simeq T_L \sim M_2$,
a $B-L$ asymmetry is generated from the $N_2$ decays.
In a second {\em wash-out stage}, at $T\sim M_1$,
inverse processes involving the lightest RH neutrinos, the $N_1$'s, become effective
and wash-out the asymmetry to some level.

In the {\em production stage}, since we assume
$10^{12}\,{\rm GeV} \gg M_2 \gg 10^{9}\,{\rm GeV}$,
the $B-L$ asymmetry is generated from the $N_2$-decays
in the so called two-flavour regime \cite{flavoreffects1,nardi1,zeno}.
In this regime the $\tau$-Yukawa interactions are fast enough to  break the coherent evolution
of the tauon component of the lepton quantum states between a decay and the subsequent
inverse decay and light flavour effects have to be taken into account in the calculation
of the final asymmetry.
On the other hand the evolution of the  muon and of
the electron components superposition  is still coherent.

If we indicate with $|\ell_{2}\rangle$ the quantum state describing the leptons produced by
$N_2$-decays, we can define the flavour branching ratios giving the
probability $P_{2\alpha}$ that $|\ell_{2}\rangle$ is measured in a flavour eigenstate $|\ell_{\a}\rangle$
as $P_{2\alpha} \equiv |\langle \ell_{\alpha}|\ell_{2}\rangle |^2$.
Analogously, indicating with $|\bar{\ell}_{2}'\rangle$ the quantum state describing the anti-leptons
produced by $N_2$-decays, we can define the anti-flavour branching ratios as
$\bar{P}_{2\alpha} \equiv|\langle\bar{\ell}_{\alpha} |\bar{\ell}'_{2}\rangle |^2$.
The tree level contribution is simply given by the average
$P^0_{2\a}=(P_{2\alpha}+\bar{P}_{2\alpha})/2$.
The total decay width of the $N_2$'s can be expressed in terms of the Dirac mass matrix as
\be
\widetilde{\G}_2 = {M_2\over 8\,\pi\,v^2}\,(m^{\dagger}_D\,m_D)_{22}
\ee
and is given by the sum $\widetilde{\G}_2=\G_2+\bar{\Gamma}_2$
of the total decay rate into leptons and of the total decay rate into
anti-leptons respectively.
The flavoured decay widths are given by
\be
\widetilde{\G}_{2\a} = {M_2\over 8\,\pi\,v^2}\,|m_{D\a 2}|^2 \, ,
\ee
and can be also expressed as a sum, $\widetilde{\G}_{2\a}=\G_{2\a}+\bar{\Gamma}_{2\a}$,
of the flavoured decay rate into leptons and of the flavoured total decay rate into
anti-leptons respectively.

Notice that the branching ratios can then be expressed in terms of the rates as
$P_{2\a}=\G_{2\a}/\G_2$ and  $\bar{P}_{2\a}=\bar{\G}_{2\a}/\bar{\G}_2$.
The flavoured  $C\!P$ asymmetries for the $N_2$-decays
into $\alpha$-leptons ($\alpha=e,\mu,\tau$) are then defined as
\be
\ve_{2\a}\equiv  -\,{\G_{2\alpha}-\overline{\G}_{2\alpha}
\over \G_{2}+\overline{\G}_{2}} \, ,
\ee
while the total  $C\!P$ asymmetries as
\footnote{Notice that we define the total and flavoured $C\!P$ asymmetries with a sign convention
in such a way that they have the same sign respectively of the produced  $B-L$ and $\D_{\alpha}$
asymmetries rather then of the $L$ and $L_{\alpha}$ asymmetries.}
\be
\ve_2\equiv  -\,{\G_2-\bar{\G}_2\over \G_2+\bar{\G}_2} =\sum_\a \ve_{2\a} \,.
\ee
The three flavoured $C\!P$ asymmetries can be calculated using \cite{crv}
\be\label{eps2a}
\ve_{2\a}=
\frac{3}{16 \p (h^{\dag}h)_{22}} \sum_{j\neq 2} \left\{ {\rm Im}\left[h_{\a 2}^{\star}
h_{\a j}(h^{\dag}h)_{2 j}\right] \frac{\x(x_j)}{\sqrt{x_j}}+
\frac{2}{3(x_j-1)}{\rm Im}
\left[h_{\a 2}^{\star}h_{\a j}(h^{\dag}h)_{j 2}\right]\right\} \, ,
\ee
where $x_j\equiv (M_j/M_2)^2$ and
\be\label{xi}
\xi(x)= {2\over 3}\,x\,
\left[(1+x)\,\ln\left({1+x\over x}\right)-{2-x\over 1-x}\right] \, .
\ee
The tree-level branching ratios can then be expressed as
\be
P^0_{2\a} = {\widetilde{\G}_{2\a}\over \widetilde{\Gamma}_2} + {\cal O}(\ve^2)
\simeq {|m_{D\a 2}|^2 \over (m^{\dagger}_D\,m_D)_{22} } \, .
\ee
Defining $\Delta P_{2\a}\equiv P_{2\a}-\bar{P}_{2\a}$,
it will prove useful to notice that the flavoured asymmetries can be
decomposed as the sum of two terms
\footnote{The derivation is simple and can be helpful to understand
later on phantom leptogenesis. If we write $P_{2\a}=P^0_{2\a}+\D P_{2\a}/2$
and $P_{2\a}=P^0_{2\a}-\D P_{2\a}/2$, one has
\[
\ve_{2\a}= -\,{P_{2\a}\,\G_{2}- \bar{P}_{2\a}\,\overline{\G}_{2}
\over \G_{2}+\overline{\G}_{2}}= P^0_{2\a}\,\ve_{2\a}-{\D P_{2\a}\over 2} \,.
\]
Notice that we are correcting a wrong sign in  Ref. \cite{flavorlep}.}
\cite{nardi1},
\be\label{eps2abis}
\ve_{2\a}=P^{0}_{2\a}\,\ve_2 - {\Delta P_{2\a} \over 2} \, ,
\ee
where the first term is
due to an imbalance between the total number of produced leptons and anti-leptons
and is therefore proportional to the total $C\!P$ asymmetry,
while the second originates from a different flavour composition of the
lepton quantum states  with respect to the $C\!P$ conjugated
anti-leptons quantum states.

Sphaleron processes conserve the flavoured asymmetries
$\D_{\a}\equiv B/3-L_{\a}$ ($\a=e,\m,\t $). Therefore, the Boltzmann
equations are particularly simple in terms of these quantities \cite{bcst}.
In the two-flavour regime the electron and the muon components of $|{\ell}_2\rangle$
evolve coherently  and the wash-out from inverse processes producing the $N_2$'s
acts then on the sum $N_{\D_{\g}}\equiv N_{\D_e} + N_{\D_{\m}}$. Therefore, it is convenient to
define correspondingly $P^0_{2\gamma}\equiv P^0_{2e}+P^0_{2\mu}$ and
$\ve_{2\gamma}\equiv \ve_{2e}+\ve_{2\mu}$. More generally, any quantity with a subscript
`$\g$' has to be meant as the sum of the same quantity calculated for the electron and
for the muon flavour component.

The asymmetry produced by the lightest and by the heaviest RH neutrino decays is negligible
since their $C\!P$ asymmetries are highly suppressed with the assumed mass pattern.
The set of classic kinetic equations reduces then to
a very simple one describing the asymmetry generated by the $N_2$-decays,
 \bea\label{flke}
{dN_{N_2}\over dz_2} & = & -D_2\,(N_{N_2}-N_{N_2}^{\rm eq}) \, ,\\
{dN_{\D_{\g}}\over dz_2} & = &
\ve_{2\g}\,D_2\,(N_{N_2}-N_{N_2}^{\rm eq})-
P_{2\g}^{0}\,W_2\,\sum_{\a=\g,\t}\,C_{\g\a}^{(2)}\,N_{\D_{\a}} \, ,\\
{dN_{\D_{\t}}\over dz_2} & = &
\ve_{2\t}\,D_2\,(N_{N_2}-N_{N_2}^{\rm eq})-
P_{2\t}^{0}\,W_2\,\sum_{\a=\g,\t}\,C_{\t\a}^{(2)}\,N_{\D_{\a}}  \, .
\eea
where $z_2 \equiv M_2/T$. The total $B-L$ asymmetry can then be calculated as
$N_{B-L}= N_{\D_{\t}}+N_{\D_{\g}}$.
The equilibrium abundances are given by
$N_{N_2}^{\rm eq}=z_2^2\,{\cal K}_2(z_2)/2$, where we indicated with
${\cal K}_i(z_2)$ the modified Bessel functions.
Introducing the total decay parameter $K_2\equiv \widetilde{\G}_{2}(T=0)/H(T=M_2)$,
the decay term $D_2$ can be expressed as
\be
D_2(z_2) \equiv {\widetilde{\G}_{2}\over H\,z}=K_2\,z_2\,
\left\langle {1\over\gamma} \right\rangle   \, ,
\ee
where  $\langle 1/\gamma \rangle(z_2)$ is
the thermally averaged dilation factor and is given by the
ratios ${\cal K}_1(z_2)/{\cal K}_2(z_2)$. Finally,
the inverse decays wash-out term is given by
\be\label{WID}
W_2(z_2) =
{1\over 4}\,K_2\,{\cal K}_1(z_2)\,z_2^3 \, .
\ee
The total decay parameter $K_2$ is related to the Dirac mass matrix by
\be
K_2={\mtt\over m_{\star}} \, ,
\hspace{10mm}
{\rm where}
\hspace{10mm}
\mtt\equiv{(m_D^{\dagger}\,m_D)_{22} \over M_2}
\ee
is the effective neutrino mass \cite{plumacher} and
$m_{\star}$ is equilibrium neutrino mass defined by \cite{orloff,annals}
\begin{equation}\label{d}
m_{\star}\equiv
{16\, \pi^{5/2}\,\sqrt{g_*} \over 3\,\sqrt{5}}\,
{v^2 \over M_{\rm Pl}}
\simeq 1.08\times 10^{-3}\,{\rm eV}.
\end{equation}
It will also prove convenient to introduce the flavoured effective neutrino masses
$\widetilde{m}_{2\a} \equiv P^0_{2\a}\,\mtt$ and correspondingly
the flavoured decay parameters $K_{2\a}\equiv P^0_{2\a}\,K_2 = \widetilde{m}_{2\a}/m_{\star} $,
so that $\sum_{\alpha} \widetilde{m}_{2\a}=\mtt$ and $\sum_{\alpha} K_{2\a}=K_2$.

The flavour coupling matrix $C$  \cite{bcst,spectator,racker,Antusch:2006cw,nardi1,flavorlep}  relates the asymmetries
stored in the lepton doublets and in the Higgs bosons to the
$\D_{\a}$'s. It is therefore the sum of two contributions,
\be
C_{\a\b}=C^{\ell}_{\a\b}+C^{H}_{\a\b} \, ,
\ee
the first one connecting the asymmetry in the lepton doublets and
the second connecting the asymmetry in the Higgs bosons.
Flavour dynamics couple because the generation of a leptonic asymmetry into lepton
doublets from $N_i$ decays is necessarily accompanied by a generation of a hypercharge asymmetry
into the Higgs bosons and of a baryonic asymmetry into quarks
via sphaleron processes. The asymmetry generated into the lepton doublets
is  moreover also redistributed to right handed charged particles.
The wash-out of a specific flavour asymmetry is then influenced by the
dynamics of the asymmetries stored in the other flavours because
they are linked primarily through the asymmetry into the Higgs doublets
and secondarily through the asymmetry into quarks.

The condition of chemical equilibrium gives a constraint on the chemical potential
(hence number density asymmetry) of each such species. Solving for all constraints
one obtains the $C_{\alpha\beta}$ explicitly.
If we indicate with $C^{(2)}$ the coupling matrix in the
two-flavour regime, the two contributions to the flavour coupling matrix are given by
\be C^{l(2)}=\left(\begin{array}{ccc}
417/589 & -120/589 \\ -30/589 & 390/589 \end{array}\right) \, \hspace{4mm} \mbox{\rm and}
\hspace{4mm}
C^{h(2)}=\left(\begin{array}{ccc}
164/589 & 224/589 \\
164/589 & 224/589
\end{array}\right) \, ,
\ee
and summing one obtains
\be
C^{(2)} \equiv
\left(\begin{array}{ccc}
C^{(2)}_{\g\g} & C^{(2)}_{\g\t}  \\  C^{(2)}_{\t\g} & C^{(2)}_{\t\t}
\end{array}\right) =
\left(\begin{array}{ccc}
581/589 & 104/589 \\ 194/589 & 614/589 \end{array}\right) \, .
\ee
A traditional calculation, where flavour coupling is neglected,
corresponds to approximating the $C$-matrix by the identity matrix. In this
case the evolution of the two flavour asymmetries proceeds uncoupled
and they can be easily worked out in an integral form \cite{kt,annals,flavorlep},
\be\label{solint}
N_{\D\a}(z_2)=N_{\D\a}^{\rm in}\,
e^{-P_{2\a}^0\,\int_{z_2^{\rm in}}^{z_2}\,dz_{2}'\,W_2(z_2')}
+\ve_{2\a}\,\k(z_2;K_{2\a}) \,  ,
\ee
where the efficiency factors are given by
\be\label{ef}
\k(z_2;K_{2\a})=-\int_{z_2^{\rm in}}^{z_2}\,dz_{2}'\,{dN_{N_i}\over dz_2'}\,
e^{-P_{2\a}^0\,\int_{z_2'}^z\,dz_{2}''\,W_2(z_2'')} \,.
\ee
We will neglect the first term due the presence of possible initial flavour asymmetries
and assume $z_2^{\rm in}\ll 1$.
The efficiency factors and therefore the asymmetries get frozen to 
a particular value of the temperature given by $T_{L\a}=M_2/z_B(K_{2\a})$,
where \cite{beyond}
\be
z_{B}(K_{2\a}) \simeq 2+4\,K_{2\a}^{0.13}\,e^{-{2.5\over K_{2\a}}}={\cal O}(1\div 10) \, .
\ee
Defining $T_L\equiv {\rm min}(T_{L\t},T_{L\g})$,
the total final $B-L$ asymmetry at $T_L $ is then given by
\be\label{solution}
N_{B-L}^{T\sim T_L} \simeq \ve_{2\g}\,\kappa(K_{2\g})+ \ve_{2\tau}\,\kappa(K_{2\tau}) \, .
\ee
Assuming an initial thermal $N_2$-abundance,
the final efficiency factors $\k(K_{2\a})\equiv \k(z_2=\infty,K_{2\a})$
are given approximately by \cite{flavorlep}
\be
\k(K_{2\a})\simeq \frac{2}{K_{2\a} \,
z_{\rm B}(K_{2\a})}\left[1-{\rm exp}\left(-\frac{1}{2} K_{2\a}\, z_{\rm B}(K_{2\a})\right)\right]\, .
\ee
On the other hand, in the case of vanishing initial abundances
\footnote{These analytical expressions reproduce very well the numerical
results found in \cite{Antusch:2006cw}. The difference is at most $30\%$
around $K_{2\a}\simeq 1$ and much smaller than $10\%$ elsewhere.}
, the efficiency factors
are the sum of two different contributions, a negative and a positive one,
\be
\k_{2\a}^{\rm f}
=\k_{-}^{\rm f}(K_2,P_{2\a}^{0})+
 \k_{+}^{\rm f}(K_2,P_{2\a}^{0}) \, .
\ee
The negative contribution arises from a first stage where
$N_{N_2}\leq N_{N_2}^{\rm eq}$, for $z_2\leq z_2^{\rm eq}$,
and is given approximately by
\be\label{k-}
\k_{-}^{\rm f}(K_2,P_{2\a}^{0})\simeq
-{2\over P_{2\a}^{0}}\ e^{-{3\,\pi\,K_{2\a} \over 8}}
\left(e^{{P_{2\a}^{0}\over 2}\,N_{N_2}(z_{\rm eq})} - 1 \right) \, .
\ee
The $N_2$-abundance at $z_2^{\rm eq}$ is well approximated by the expression
\begin{equation}\label{nka}
N_{N_2}(z_2^{\rm eq}) \simeq \overline{N}(K_2)\equiv
{N(K_2)\over\left(1 + \sqrt{N(K_2)}\right)^2}\, ,
\end{equation}
that interpolates between the limit $K_2\gg 1$, where $z_2^{\rm eq}\ll 1$ and
$N_{\rm N_2}(z_2^{\rm eq})=1$, and the limit $K_2\ll 1$, where
$z_2^{\rm eq}\gg 1$ and $N_{N_2}(z_2^{\rm eq})=N(K_2)\equiv 3\p K_2/4$.
The positive contribution arises from a second stage when
$N_{N_2}\geq N_{N_2}^{\rm eq}$, for $z_2\geq z_2^{\rm eq}$,
and is approximately given by
\be\label{k+}
\k_{+}^{\rm f}(K_2,P_{2\a}^{0})\simeq
{2\over z_B(K_{2\a})\,K_{2\a}}
\left(1-e^{-{K_{2\a}\,z_B(K_{2\a})\,N_{N_2}(z_{\rm eq})\over 2}}\right) \, .
\ee
If flavour coupling is taken into account, we can still solve analytically eqs.~(\ref{flke})
performing the following change of variables
\be
\left(\begin{array}{c}
 N_{\D_{\g'}}  \\
N_{\D_{\t'}}
\end{array}\right) =  U\,
\left(\begin{array}{c}
N_{\D_{\g}}  \\
N_{\D_{\t}}
\end{array}\right) \, , \hspace{5mm} \mbox{\rm where} \hspace{5mm}
U\equiv \left(\begin{array}{cc}
U_{\g'\g} & U_{\g'\t}  \\
U_{\t'\g} & U_{\t'\t}
\end{array}\right)
\ee
is the matrix that diagonalizes
\be
P^0_2 \equiv
\left(\begin{array}{cc}
P^0_{2\g}\,C_{\g\g}^{(2)} & P^0_{2\g}\,C_{\g\t}^{(2)}  \\
P^0_{2\tau}\,C_{\t\g}^{(2)} & P^0_{2\tau}\,C_{\t\t}^{(2)}
\end{array}\right) \, ,
\ee
i.e. $U\,P^0_{2}\,U^{-1} ={\rm diag}(P^0_{2\g'},P^0_{2\t'})$.
In these new variables the two kinetic
equations for the flavoured asymmetries decouple,
 \bea\label{flke2}
{dN_{\D_{\g'}}\over dz_2} & = &
\ve_{2\g'}\,D_2\,(N_{N_2}-N_{N_2}^{\rm eq})-P^0_{2\g'}\,W_2\,N_{\D_{\g'}} \, \\
{dN_{\D_{\t'}}\over dz_2} & = &
\ve_{2\t'}\,D_2\,(N_{N_2}-N_{N_2}^{\rm eq})-P^0_{2\t'}\,W_2\,N_{\D_{\t'}}  \, ,
\eea
where we defined
\be
\left(\begin{array}{c}
 \ve_{2\g'}  \\
\ve_{2\t'}
\end{array}\right) \equiv  U\,
\left(\begin{array}{c}
\ve_{2\g}  \\
\ve_{2\t}
\end{array}\right) \, .
\ee
The solutions for the two $N_{\D_{\a'}}$ are then still given by eq.~(\ref{solint})
where, however, now the `unprimed'
quantities have to be replaced with the `primed' quantities and therefore
explicitly one has
\bea\label{solution2}
N_{\D_{\g'}}^{T\sim T_L}  & \simeq &
\ve_{2\g'}\,\kappa(K_{2\g'}) \, , \\ \nonumber
N_{\D_{\t'}}^{T\sim T_L}  & \simeq &
\ve_{2\tau'}\,\kappa(K_{2\tau'}) \, .
\eea
Notice that the $B-L$ asymmetry at $T\sim T_L$ is still given by $N_{B-L}^{T\sim T_L}=
N_{\D_{\g}}^{T\sim T_L}+N_{\D_\t}^{T\sim T_L}$.
The two $N_{\D_{\a}}$'s can be calculated from the two $N_{\D_{\a'}}$'s
using the inverse transformation
\be\label{solutioninv}
\left(\begin{array}{c}
 N_{\D_{\g}}^{T\sim T_L} \\
N_{\D_{\t}}^{T\sim T_L}
\end{array}\right) =  U^{-1}\,
\left(\begin{array}{c}
N_{\D_{\g'}}^{T\sim T_L}  \\
N_{\D_{\t'}}^{T\sim T_L}
\end{array}\right) \, , \hspace{5mm} \mbox{\rm where} \hspace{5mm}
U^{-1}\equiv \left(\begin{array}{cc}
U^{-1}_{\g\g'} & U^{-1}_{\g\t'}  \\
U^{-1}_{\t\g'} & U^{-1}_{\t\t'}
\end{array}\right) \, .
\ee
To study the impact of flavour coupling on the final asymmetry, we can calculate the ratio
\be\label{r}
R \equiv \left|{N_{B-L}\over \left.N_{B-L}\right|_{C=I}}\right|
\ee
between the asymmetry calculated taking into account flavour coupling,
and the asymmetry calculated neglecting flavour coupling, corresponding to the assumption $C=I$.
If we want first to calculate the value of $R$ at the production stage,
we have to express $N_{B-L}^{T\sim T_L}$
in terms of the `unprimed' quantities in eq.~(\ref{solution2}).
This is quite easy for the $K_{2\a'}$, since one has simply to find
the eigenvalues of the matrix $P_2^0$. Taking for simplicity the
approximation $C^{(2)}_{\g\g}\simeq C^{(2)}_{\t\t} \simeq 1$,
and remembering that $P_{2\gamma}^0+P_{2\tau}^0=1$, one obtains
\bea\label{eigenvalues}
P^0_{2\g'} & \simeq & {1\over 2}\,\left(1+\sqrt{(P^0_{2\g}-P^0_{2\t})^2+
4\,C^{(2)}_{\g\t}\,C^{(2)}_{\t\g}\,P^0_{2\g}\,P^0_{2\t}}\right) \, ,\\
P^0_{2\tau'} & \simeq & {1\over 2}\,\left(1-\sqrt{(P^0_{2\g}-P^0_{2\t})^2+
4\,C^{(2)}_{\g\t}\,C^{(2)}_{\t\g}\,P^0_{2\g}\,P^0_{2\t}}\right) .
\eea
Notice that, both for $\a=\t$ and $\a=\g$, one has
$P^0_{2\a'}\simeq P^0_{2\a}+{\cal O}(\sqrt{C^{(2)}_{\g\t}\,C^{(2)}_{\t\g}})$
if $P_{2\t}\simeq P_{2\g}^0\simeq 1/2$ and
$P^0_{2\a'}\simeq P^0_{2\a}+{\cal O}(C^{(2)}_{\g\t}\,C^{(2)}_{\t\g})$
if $P_{2\t}\ll P_{2\g}^0$ or vice-versa.
Considering moreover  that, if $K_{2\a} \gg 1$, one has approximately
$\k(K_{2\a})\sim 1/K_{2\a}^{1.2}$,
one can write
\bea\label{solution3}
N_{\D_{\g'}}^{T\sim T_L}  & \simeq &
\ve_{2\g'}\,\kappa(K_{2\g}) \, , \\ \nonumber
N_{\D_{\t'}}^{T\sim T_L}  & \simeq &
\ve_{2\tau'}\,\kappa(K_{2\tau}) \, .
\eea
We have now to consider the effect of flavour coupling encoded in
the primed $C\!P$ asymmetries. If these are re-expressed
in terms of the unprimed $C\!P$ asymmetries we can obtain
explicitly the flavour composition
of the asymmetry generated at $T\simeq T_L$ plugging  eqs.~(\ref{solution3})
into eqs.~(\ref{solutioninv}),
\bea \label{Dg}
N_{\D_{\g}}^{T\sim T_L} & = &
 U^{-1}_{\g\g'}\left[U_{\g'\g}\,\ve_{2\g}+U_{\g'\t}\,\ve_{2\t}\right]\,\kappa(K_{2\g})
+U^{-1}_{\g\t'}\left[U_{\t'\g}\,\ve_{2\g}+U_{\t'\t}\,\ve_{2\t}\right]\,\kappa(K_{2\t}) \, , \\ \label{Dtau}
N_{\D_{\t}}^{T\sim T_L} & = &
U^{-1}_{\t\g'}\left[U_{\g'\g}\,\ve_{2\g}+U_{\g'\t}\,\ve_{2\t}\right]\,\kappa(K_{2\g})
+U^{-1}_{\t\t'}\left[U_{\t'\g}\,\ve_{2\g}+U_{\t'\t}\,\ve_{2\t}\right]\,\kappa(K_{2\t}) \, , \\
N_{B-L}^{T\sim T_L}  & = & N_{\D_{\g}}^{T\sim T_L} + N_{\D_\t}^{T\sim T_L} \, .  \label{flas}
\eea
We can distinguish two different cases. The first one is for $P^0_{2\t}\simeq P^0_{2\g}\simeq 1/2$,
implying $K_{2\t}=K_{2\g}=K_2/2$ and therefore $\kappa(K_{2\g})=\kappa(K_{2\t})=\kappa(K_2/2)$.
In this situation one can see immediately that
\be
N_{\D_{\g}}^{T\sim T_L} \simeq \ve_{2\g}\,\kappa(K_2/2) \, , \hspace{5mm}
\mbox{\rm and} \hspace{5mm} N_{\D_{\t}}^{T\sim T_L} \simeq \ve_{2\t}\,\kappa(K_2/2) \, .
\ee
Therefore, barring the case $\ve_{2\g}=-\ve_{2\t}$,
one has not only $N_{B-L}^{T\sim T_L}\simeq \left.N_{B-L}^{T\sim T_L}\right|_{C=I}$,
 implying $R^{T\sim T_L}=1$, but even that the flavour composition is the same
 compared to a usual calculation  where flavour coupling is neglected. However,
 if $\ve_{2\g}=-\ve_{2\t}$, a more careful treatment is necessary. From the
 eqs.~(\ref{eigenvalues}) one finds $P^0_{2\g'}=(1+\sqrt{C^{(2)}_{\g\t}\,C^{(2)}_{\t\g}})/2\neq
 P^0_{2\t'}=(1-\sqrt{C^{(2)}_{\g\t}\,C^{(2)}_{\t\g}})/2$. This difference induced by the off-diagonal
 terms of the $C^{(2)}$ matrix prevents an exact cancelation or at least it changes the condition
 where it is realized, an effect that occurs also within $N_1$ leptogenesis \cite{abada}.

Let us now see what happens on the other hand when either $P^0_{2\t}$ or  $P^0_{2\g}$
is much smaller than the other. This situation has not to be regarded as fine tuned,
since it occurs quite naturally for a random choice of the parameters.
At the first order in the $C^{(2)}$ off-diagonal terms, one has
\be
U \simeq \left(\begin{array}{cc}
1 &  C^{(2)}_{\g\t}\,{P^0_{2\g}\over P^0_{2\g}-P^0_{2\t}} \\
C^{(2)}_{\t\g} {P^0_{2\t}\over P^0_{2\t}-P^0_{2\g}} & 1
\end{array}\right)
\, ,
\,\,\, U^{-1} \simeq \left(\begin{array}{cc}
1 &  -C^{(2)}_{\g\t}\,{P^0_{2\g}\over P^0_{2\g}-P^0_{2\t}} \\
-C^{(2)}_{\t\g} {P^0_{2\t}\over P^0_{2\t}-P^0_{2\g}} & 1
\end{array}\right)
\, .
\ee
Let us for definiteness assume that $P^0_{2\t}\ll P^0_{2\g}$ and that $K_2 \gg 1$ (this second
condition also occurs for natural choices of the parameters).
In this case one has necessarily $\k(K_{2\tau})\gg \k(K_{2\g})$.
We can therefore specify eqs.~(\ref{flas})
writing approximately for the flavour asymmetries in the two flavours,
\bea\label{flasspec}
N_{\D_{\g}}^{T\sim T_L} & \simeq & \ve_{2\g}\,\k(K_{2\g}) - C^{(2)}_{\g\t}\,\ve_{2\t}\,\k(K_{2\t}) \, ,
\\ \label{flasspec2}
N_{\D_{\t}}^{T\sim T_L} & \simeq & \ve_{2\t}\,\k(K_{2\t}) \, ,
\eea
where we neglected all terms containing products either of two off-diagonal terms of $C^{(2)}$,
or of one off-diagonal term times $\k(K_{2\g})$.
We can therefore see that the total asymmetry cannot
differ much from the standard calculation,
\be
N_{B-L}^{T\sim T_L}\simeq \ve_{2\g}\,\k(K_{2\g}) + \ve_{2\t}\,\k(K_{2\t})
- C^{(2)}_{\g\t}\,\ve_{2\t}\,\k(K_{2\t}) \, ,
\ee
implying
\be\label{RTL}
R^{T\sim T_L} \simeq \left|1 - C^{(2)}_{\g\t}\,
{\ve_{2\t}\,\k(K_{2\t})\over \ve_{2\g}\,\k(K_{2\g}) + \ve_{2\t}\,\k(K_{2\t})}\right| \, .
\ee
This holds because the dominant contribution comes from the tauonic flavour asymmetry
that is not changed at first order. Notice by the way that since $C^{(2)}_{\g\t}>0$
and necessarily  $\ve_{2\t}>0$, the effect of flavour coupling even produces
a reduction of the total asymmetry at $T\sim T_L$
\footnote{This result differs from the one of \cite{abada} where, within $N_1$ leptogenesis,
the authors find an enhancement instead of a reduction. This is simply
explained by the fact that we are also accounting for the Higgs asymmetry that determines
the (correct) positive sign  for $C^{(2)}_{\g\t}$.}.

On the other hand the asymmetry in the sub-dominant flavour $\g$ can be
greatly enhanced since the quantity
\be
R_{\D_\g}^{T\sim T_L}\equiv \left|{N_{\D_{\g}}^{T\sim T_L}\over
\left.N_{\D_{\g}}^{T\sim T_L}\right|_{C=I}}\right| \simeq
\left|1 - C^{(2)}_{\g\t}\,{\ve_{2\t}\,\k(K_{2\t})\over \ve_{2\g}\,\k(K_{2\g}) } \right|
\ee
can be in general much higher than unity. In this respect it is important
to notice that the assumption $P^0_{2\t}\ll P^0_{2\g}$ does not necessarily imply
$\ve_{2\t} \ll \ve_{2\g}$ since $\ve_{2\a}\lesssim 10^{-6}\,(M_2/10^{10}\,{\rm GeV})\,\sqrt{P^0_{2\a}}$.
Notice also that if vice versa $P^0_{2\g}\ll P^0_{2\t}$, then the $\t$
flavour asymmetry is sub-dominant and can be strongly enhanced.

There is a simple physical interpretation to the enhancement of the sub-dominant flavoured
asymmetry. This can be given in terms of the effect of tau flavour  coupling on the final
$\g$ asymmetry that is described by the off-diagonal terms of the $C^{(2)}$ matrix.
The dominant contribution to these terms comes  from  the Higgs asymmetry produced
in $N_2 \ra l_{\alpha}+\phi^{\dagger}$ decays. Let us still
assume for definiteness that  $P^0_{2\t}\ll P^0_{2\g}$ and that $K_2\gg 1$.
This implies that the $\g$ asymmetry is efficiently washed-out and there is
a substantial equilibrium between decays and inverse processes.

On the other hand the $\t$ asymmetry is weakly washed-out and for simplicity
we can think to the extreme case when is not washed-out at all (true for $K_{2\t}\ll 1$).
An excess of tau over $\g$  asymmetry results in an excess of
Higgs over $\g$ asymmetry.
This excess Higgs asymmetry increases the inverse decays of ${\ell}_\g$ over the
$\bar{\ell}_\g$ states (or vice versa, depending on its sign) and
`soaks up' either more particle or more anti-particle states
generating an imbalance.
Hence one can have $R_{\D\g}^{T\sim T_L}\gg 1$ thanks to the
dominant effect of the extra inverse decay processes
that `switch on' when $C \neq I$.

This effect had been already discussed within $N_1$-dominated leptogenesis \cite{abada}.
Our results, for the asymmetry at the production stage, are qualitatively similar though
we also took into account the dominant contribution to flavour coupling
coming from the Higgs asymmetry and we
solved analytically the kinetic equations including flavour
coupling without any approximation. As we already noticed, quantitatively,
the account of the Higgs asymmetry produces important effects. For instance, when the
Higgs asymmetry is included, the results are quite symmetric under the interchange of
$P^0_{2\g}$ and $P^0_{2\t}$ since the total matrix
$C^{(2)}$ is much more symmetrical than $C^{l(2)}$.

There is however a much more important difference in this respect between $N_2$-dominated
 and $N_1$-dominated leptogenesis. While in the latter case a strong
enhancement of the sub-dominant flavoured asymmetry does not translate into a strong
enhancement of the final asymmetry, in the case of the $N_2$-dominated scenario this becomes possible,
thanks to the presence of the additional stage of lightest RH neutrino wash-out,
as we discuss in the next section.

\section{Three flavour projection and the $N_1$ wash-out stage}

At $T\sim 10^{9}\,{\rm GeV}$ the muon Yukawa interactions equilibrate as well.
They are able to break the residual coherence of the superposition of the muon and
electron components of the quantum states $|{\ell}_2\rangle$ and $|\bar{\ell}'_2\rangle$ .
Consequently, the `$\g$' asymmetry becomes
an incoherent mixture of an electron and a muon component \cite{decoherence2} and
the three-flavour regime holds \cite{flavoreffects1,nardi1}.

Therefore, for temperatures $T'$ such that $10^{9}\,{\rm GeV}\gg T' \gg M_1$, one has a situation where
the asymmetry in the tau flavour is still given by the frozen value produced at $T\sim T_L$ (cf. eq.~(\ref{Dtau})),
whereas the asymmetries in the electron and in the muon flavours have to be calculated splitting the
$\g$-asymmetry produced at $T\sim T_L$ (cf. eq.~(\ref{Dg})) and the result is
\bea\label{Demu}
N_{\D_{\d}}(T') & =  & p_{\d}+{P^0_{2\d}\over P^0_{2\g}} \,N_{\D_{\g}}^{T\sim T_L}\, , \hspace{10mm} (\d=e,\m )
\eea
where the ``phantom terms'' $p_{e}$ and $p_{\m}$, for an initial thermal $N_2$-abundance $N_{N_2}^{\rm in}$,
are given by
\bea
\hspace{15mm} p_{\d} & = &  \left(\ve_{2\d}- {P^0_{2\d}\over P^0_{2\g}}\,{\ve_{2\g}}\right)\,\,N_{N_2}^{\rm in} \, ,\hspace{10mm} (\d=e,\m )
\eea
and one can easily check that $p_e+p_\m=0$. Notice that, because of the presence of the phantom terms,
the electron and the muon components are not just proportional to $\ve_{2\g}$.

Let us  show in detail how the result eq.~(\ref{Demu}) and the expression for the
phantom terms can be derived. The derivation is simplified if one considers
the $\D_{\d}$ asymmetry as the result of two separate stages: first
an asymmetry $N_{L_\d}^\star$ ends up, at the break of coherence, into the $\d$ lepton doublets
and then it is flavour redistributed and sphaleron-converted in a way that $N_{\D_\d}=-N_{L_\d}^\star$.
Actually part of the $N_{L_\d}$
asymmetry gets redistributed and sphaleron-converted immediately  after having been produced.
However, in our simplified procedure, the notations is greatly simplified
and the derivation made more transparent but the final result does not change, since flavour redistribution
and sphalerons conserve the $\D_{\d}$ asymmetries.

After these premises, we can say that the asymmetry in the $\d$ lepton doublets
at the break of coherence is simply given by
\be
N_{L_{\d}}^{\star}=f_{2\d}\,N_{\ell_{\g}}^{T\sim T_L}-\bar{f}_{2\d}\,N_{\bar{\ell}_{\gamma}}^{T\sim T_L} \, ,
\ee
where $f_{2\d} \equiv |\langle \ell_{\d}|\ell_{2\g}\rangle |^2 = {P_{2\d}/P_{2\g}}$
and $\bar{f}_{2\d} \equiv |\langle \ell_{\d}|\bar{\ell}'_{2\g}\rangle |^2 = \bar{P}_{2\d}/\bar{P}_{2\g}$.
With some easy passages one can then write
\bea
N_{L_{\d}}^{\star}
& = & {1\over 2}\left(f_{2\d}-\bar{f}_{2\d}\right)\,\left(N_{\ell_{\g}}^{T\sim T_L}+N_{\bar{\ell}_{\gamma}}
^{T\sim T_L}\right) \\
& + & {1\over 2}\left(f_{2\d}+\bar{f}_{2\d}\right)\,N_{L_\g}^{T\sim T_L} \\
& = & - p_\d + {1\over 2}\,\left({f_{2\d}+\bar{f}_{2\d}}\right)\,N_{L_\g}^{T\sim T_L} \, ,
\eea
where in the last expression we introduced the phantom term
\be
p_\d= - {1\over 2}\left(f_{2\d}-\bar{f}_{2\d}\right)\,\left(N_{\ell_{\g}}^{T\sim T_L}+N_{\bar{\ell}_{\gamma}}^{T\sim T_L}\right) \, .
\ee
Considering now that
$N_{\ell_{\g}}^{T\sim T_L}+N_{\bar{\ell}_{\gamma}}^{T\sim T_L}\simeq P^0_{2\g}\,N_{N_2}^{\rm in}$
and that, using first $f_{2\d} = {P_{2\d}/P_{2\g}}$
and  $\bar{f}_{2\d} = \bar{P}_{2\d}/\bar{P}_{2\g}$ and then the eq.~(\ref{eps2abis}),
one has
\bea
{1\over 2}\left(f_{2\d}-\bar{f}_{2\d}\right)\,P^0_{2\g} & \simeq &
- \left(\ve_{2\d}-{P^0_{2\d}\over P^0_{2\gamma}}\,\ve_{2\g}\right)  \\
{1\over 2}\left(f_{2\d}+\bar{f}_{2\d}\right) & = & {P^0_{2\d}\over P^0_{2\gamma}} \, ,
\eea
one finally finds
\be
N_{L_\d}^{\star}= - p_\d +{P^0_{2\d}\over P^0_{2\g}}  \,N_{L_\g}^{T\sim T_L} \, ,
\ee
where the phantom terms can be expressed in terms of the $C\!P$ asymmetries as
\be
p_\d = \left(\ve_{2\d}-{P^0_{2\d}\over P^0_{2\gamma}}\,\ve_{2\g}\right)\,N_{N_2}^{\rm in}  \, .
\ee
As a last step one has finally to take into account flavour redistribution
and sphaleron conversion so that the eq.~(\ref{Demu}) follows.

 The phantom terms  originate
from the second contribution in eq.~(\ref{eps2abis}) to the flavoured $C\!P$ asymmetries.
One can see indeed that if $\Delta P_{2e}= \Delta P_{2\m} = 0$, then $p_{e}=p_{\m}=0$.
On the other hand, these terms do not vanish if the leptons and the anti-leptons
produced by the decays have a different flavour composition, such that at least one $\Delta P_{2\d}\neq 0$,
even when $\ve_{2\g}=0$. In this particular case one can indeed see that
$p_{e}= \ve_{2e}=-\ve_{2\mu}=-p_{\m}$ \, .

It should be noticed that, remarkably, the phantom terms are not washed-out at the production.
This happens because in this stage the $e$ and $\m$ components of the leptons and anti-leptons
quantum states are still
in a coherent superposition. The phantom terms originate from the components of the
electron and muon asymmetries dependant only on differences between the
flavour compositions of the leptonic quantum states ${\ell_{2\g}}$ and
anti-lepton quantum states ${\bar{\ell}'_{2\g}}$. These cannot be
washed-out by the $N_2$ inverse processes, which can only act to destroy
the part of the electron and muon asymmetries
proportional to $\ve_{2\g}$ itself
\footnote{The name {\em phantom} is not meant to imply that the
effect is non physical. It is simply justified by the fact that the effect arises from
terms which cancel and are therefore invisible (i.e. phantom-like) until a possible wash-out from the $N_1$
acts asymmetrically on the $e$ and $\mu$ components ($K_{1e}\neq K_{1\mu}$),
which renders the difference observable.}.

However, it should be also noticed that if one assumes
an initial vanishing $N_2$-abundance, the phantom terms vanish. This
happens because
in this case they would be produced during the $N_2$ production stage with an opposite sign
with respect to the decay stage such that an exact cancelation would occur implying a vanishing
final value
\footnote{This can be understood, for example,  in the following way. An inverse decays of a lepton
with an Higgs, corresponds to the creation either of a state orthogonal to $|{\ell_{2\g}}\rangle$,
that we indicate with $|{\ell_{2\g}^{\bot}}\rangle$,  or to $|\bar{\ell}_{2\g}'\rangle$, that we indicate
with $|\bar{\ell}_{2\g}^{'\bot}\rangle$. Their flavour composition is given
by $|{\ell}_{2\g}^{\bot}\rangle =
\langle {\ell}_\mu|{\ell}_{2\g} \rangle\,|{\ell_e}\rangle-
\langle {\ell}_e|{\ell}_{2\g} \rangle\,|{\ell}_\mu\rangle$
and by $|\bar{\ell}_{2\g}^{'\bot}\rangle =\langle \bar{\ell}_\mu|\bar{\ell}'_{2\g} \rangle\,
|\bar{\ell}_e\rangle-\langle \bar{\ell}_e|\bar{\ell}'_{2\g} \rangle\,|\bar{\ell}_\mu\rangle$.
Therefore, each inverse decay will produce, on average, an electron and a muon asymmetry given respectively
by $\Delta L_e^{id}=(f_{2\m}-\bar{f}_{2\m})/2$ and $\D L_\m^{id}=(f_{2e}-\bar{f}_{2e})/2$,
opposite to those produced by one decay. Notice that only $N_2$ inverse processes
can produce such $C\!P$ violating orthogonal states with phantom terms exactly
canceling with those in the lepton quantum states produced from decays}.
Therefore, the phantom terms seem to introduce a strong dependence on the initial conditions
in $N_2$-flavoured leptogenesis.

When finally the inverse processes involving the lightest RH neutrinos
become active at $T\sim M_1$, the wash-out from the $N_1$-decays
acts separately on the three flavour components of the total $B-L$ asymmetry \cite{vives}.

 The wash-out from the lightest RH neutrinos is more efficient than the wash-out
from the next-to-lightest RH neutrinos  since it is not balanced by any production
and it therefore acts on the whole produced asymmetry.

Taking into account the flavour coupling matrix,
the set of kinetic equations describing this stage is given by
\be\label{flkewA}
{dN_{\D_{\a}}\over dz_1}  =
-P_{1\a}^{0}\,\sum_{\b}\,C^{(3)}_{\a\b}\,W_1^{\rm ID}\,N_{\D_{\b}} \, ,
\hspace{15mm} (\a,\b=e,\m,\t)
\ee
where $z_1\equiv M_1/T$ and, more generally, all
quantities previously defined for the $N_2$'s can be also
analogously defined for the $N_1$'s. In particular
the $P_{1\a}^{0}$'s, the $K_{1\a}$'s and $W_1$
are defined analogously to  the $P_{2\a}^{0}$, to the $K_{2\a}$'s and
to $W_2$ respectively.

The flavour coupling matrices in the three-flavour regime are given by
\[C^{l(3)}=\left(\begin{array}{ccc}
151/179 & -20/179 & -20/179 \\ -25/358 & 344/537 & -14/537 \\ -25/358 & -14/537 & 344/537	
\end{array}\right) \, , \hspace{5mm}
C^{h(3)}=\left(\begin{array}{ccc}
37/179 & 52/179 & 52/179 \\
37/179 & 52/179 & 52/179 \\
37/179 & 52/179 & 52/179 \\
\end{array}\right) \, ,
\]
\[C^{(3)} \equiv
\left(\begin{array}{ccc}
C_{ee}^{(3)} & C_{e\m}^{(3)} & C_{e\t}^{(3)} \\
C_{\m e}^{(3)} & C_{\m\m}^{(3)} & C_{\m \t}^{(3)} \\
C_{\t e}^{(3)} & C_{\t\m}^{(3)} & C_{\t\t}^{(3)}	
\end{array}\right) =
\left(\begin{array}{ccc}
188/179 & 32/179 & 32/179 \\ 49/358 & 500/537 & 142/537 \\ 49/358 & 142/537 & 500/537	
\end{array}\right) \, .
\]
If flavour coupling is neglected both at the production in the
two-flavour regime (corresponding to the approximation $C^{(2)}=I$)
and in the lightest RH neutrino wash-out in the three-flavour regime
(corresponding to the approximation $C^{(3)}=I$),
the final asymmetry is then given by \cite{aspects,SO10}
\bea\label{finalasnoc}
N^{\rm f}_{B-L} & = & \sum_\alpha N_{\Delta\alpha}^f \nonumber \\
&=& \sum_{\d=e,\m} \left[p_\d+ {P^0_{2\d}\over P^0_{2\g}}\,\ve_{2 \g}\,\kappa(K_{2\g})\right]\,
e^{-{3\pi\over 8}\,K_{1 \d}} +
\ve_{2 \tau}\,\kappa(K_{2 \tau})\,e^{-{3\pi\over 8}\,K_{1 \tau}} \, .
\eea
It is interesting that, even though $K_1\gg 1$,
there can be a particular flavour  $\a$ with at the same time
$1 \simeq K_{1\a}\ll K_1 $ and a sizeable
$\ve_{2\a}= {\cal O}(10^{-5}-10^{-6})$.
In this case the final asymmetry is dominated by this particular
$\a$-flavour  contribution, avoiding the lightest RH neutrino wash-out,
and can reproduce the observed asymmetry.
Therefore, thanks to flavour effects, one can have successful
leptogenesis even for $K_1\gg 1$, something otherwise impossible in
the unflavoured regime \cite{vives,aspects,SO10}.

Let us now comment on the phantom terms $p_{\d}$ and
on the conditions for them to be dominant so that
a scenario of `phantom leptogenesis' is realized.
First of all let us importantly recall that we are assuming zero pre-existing
asymmetries. Under this assumption the phantom terms would be
present only for a non zero initial  $N_2$ abundance
while they would vanish if  an initial vanishing $N_2$ abundance is assumed.

A condition for phantom leptogenesis is then
\be
|p_{\d}|\gg
\left|{P^0_{2\d}\over P^0_{2\g}}\,\ve_{2 \g}\,\kappa(K_{2 \g})\right|
\;\; \mbox{\rm and} \;\; K_{1\d}\lesssim 1 \, ,
\ee
either for $\d=e$ or for $\d=\m$ or for both. In this situation the final asymmetry
will be dominated by that part of the electron-muon asymmetries that escape the
wash-out at the production thanks to the quantum coherence during the two flavour regime.
A first obvious condition is $p_{\d}\neq 0$. Another condition is to have
$K_{2\g}\gg 1$ since otherwise
the phantom terms are not crucial to avoid the wash-out at the production
that would be absent anyway.
Another necessary condition for the phantom leptogenesis scenario to hold
is that either $K_{1e}\lesssim 1$ or
$K_{1\m}\lesssim 1$, otherwise both the electron and the muon asymmetries, escaping the
wash-out at the production, are then later on washed-out
by the lightest RH neutrino wash-out processes. However, as we will see, this
condition is not necessary when the flavour coupling at the lightest RH neutrino wash-out stage
is also taken into account.

Conversely a condition for `non-phantom leptogenesis' relies on the following possibilities:
either an initial vanishing $N_2$ abundance,
or  $p_{\delta}\simeq 0$, or  $K_{2\g}\ll 1$, or that both $K_{1e}\gg 1$
and $K_{1\m}\gg 1$. Again this third condition seems however not to be sufficient
to avoid the appearance of phantom terms in the expression of the final asymmetry
when the flavour coupling at the lightest RH neutrino wash-out stage
is also taken into account. Therefore, it should be noticed that
the effects of flavour coupling and of phantom terms cannot be easily disentangled.
Notice that a last condition for non-phantom leptogenesis is
$\exp[-3\,\pi\,K_{1e}/8]\simeq \exp[-3\,\pi\,K_{1\m}/8]$, since in this
case the two terms would continue to cancel with each other even after
the lightest RH neutrino wash-out.  In the Appendix B we report a description of phantom leptogenesis
within a density matrix formalism \cite{preparation} arriving to the same conclusions and results.

In the following we will focus on the effects induced by flavour
coupling, also in transmitting the phantom terms from the electron and muon
flavours to the tauon flavour.

Let us now see how the eq.~(\ref{finalasnoc}) gets modified when
flavour coupling is taken into account (only) at the production.
In this case one has
\bea\label{finalascoup2}
N^{\rm f}_{B-L} & = &
N_{\D_{e}}^{T\sim T_L}\, e^{-{3\pi\over 8}\,K_{1 e}}+
N_{\D_{\m}}^{T\sim T_L}\, e^{-{3\pi\over 8}\,K_{1 \mu}}+
N_{\D_{\t}}^{T\sim T_L} \,e^{-{3\pi\over 8}\,K_{1 \tau}} \, ,
\eea
where $N_{\D_{e}}^{T\sim T_L}$, $N_{\D_{\m}}^{T\sim T_L}$ and
$N_{\D_{\t}}^{T\sim T_L}$ are given by eqs.~(\ref{flas}) and (\ref{Demu}).
In the specific case when $P^0_{2\t}\ll P^0_{2\g}$, the eqs.~(\ref{flas})
specialize into eqs.~(\ref{flasspec}) and (\ref{flasspec2}) and we can therefore write
\bea\label{finalascoup3}
N^{\rm f}_{B-L} & = &
\left(p_e+{P^0_{2e}\over P^0_{2\g}}\,\left[\ve_{2\g}\,\k(K_{2\g}) - C^{(2)}_{\g\t}\,\ve_{2\t}\,\k(K_{2\t})\right]\right)\,
e^{-{3\pi\over 8}\,K_{1 e}}+ \\ \nonumber
& & \left(p_\m+{P^0_{2e}\over P^0_{2\g}}\,
\left[\ve_{2\g}\,\k(K_{2\g}) - C^{(2)}_{\g\t}\,\ve_{2\t}\,\k(K_{2\t})\right]\right)
\, e^{-{3\pi\over 8}\,K_{1 \mu}}+ \\ \nonumber
&  & \ve_{2\tau}\, \k(K_{2\t})\,e^{-{3\pi\over 8}\,K_{1 \tau}} \, .
\eea
Let us finally also examine the changes induced
by flavour coupling in the description of the lightest
RH neutrino wash-out stage in the three-flavour regime, removing
the approximation $C^{(3)}=I$. One can see from eqs.~(\ref{flkewA}),
that the wash-out acts in a coupled way on the three-flavour
components of the asymmetry. An exact analytical solution can be obtained
applying again the same procedure as in the two flavour regime.
If we define
\be
P_1^0 \equiv
\left(\begin{array}{ccc}
P^0_{1e}\,C_{ee}^{(3)} & P^0_{1e}\,C_{e\m}^{(3)} & P^0_{1e}\,C_{e\t}^{(3)} \\
P^0_{1\m}\,C_{\m e}^{(3)} & P^0_{1\m}\,C_{\m\m}^{(3)} & P^0_{1\m}\,C_{\m \t}^{(3)} \\
P^0_{1\t}\,C_{\t e}^{(3)} & P^0_{1\t}\,C_{\t\m}^{(3)} & P^0_{1\t}\,C_{\t\t}^{(3)} 	
\end{array}\right) \, ,
\ee
the set of kinetic equations can be recast in a compact matrix form as
\be
{d\vec{N}_{\D}\over dz_1}  = - W_1\,P_1^0\, \vec{N}_{\D} \, ,
\ee
where $\vec{N}_{\D}\equiv (N_{\D_e},N_{\D_\m},N_{\D_\t})$.
If we perform the change of variables
\be\label{V}
\vec{N}_{\D''}= V\,\vec{N}_{\D} \, , \hspace{4mm} \mbox{\rm where}
\hspace{5mm}
V\equiv \left(\begin{array}{ccc}
V_{ e'' e} & V_{e''\m} & V_{e''\t} \\
V_{\m'' e} & V_{\m''\m} & V_{\m''\t} \\
V_{\t'' e} & V_{\t''\m} & V_{\t''\t}
\end{array}\right)
\ee
is the matrix that diagonalizes $P^0_1$,
i.e. $V\,P^0_{1}\,V^{-1} = P^0_{1''} \equiv {\rm diag}(P^0_{1 e''},P^0_{1\m''},P^0_{1\t''}) $
and $\vec{N}_{\D''}\equiv (N_{\D_{e''}},N_{\D_{\m''}},N_{\D_{\t''}})$,
the kinetic equations for the flavoured asymmetries decouple and can be written as
\be
{d\vec{N}_{\D''}\over dz_1}  =
- W_1\,P^0_{1''}\, \vec{N}_{\D''} \, .
\ee
The solution in the new variables is now given straightforwardly by
\be
\vec{N}_{\D''}^{\rm f} =
\left(
N_{\D_{e''}}^{T\sim T_L}\,e^{-{3\,\pi\over 8}\,K_{1e''}}, \,
N_{\D_{\m''}}^{T\sim T_L}\,e^{-{3\,\pi\over 8}\,K_{1 \m''}}, \,
N_{\D_{\t''}}^{T\sim T_L}\,e^{-{3\,\pi\over 8}\,K_{1 \t''}}\right) \, ,
\ee
where $K_{1\a''}\equiv P^0_{1\a''}\,K_1$. Applying the inverse transformation,
we can then finally obtain the final flavoured asymmetries
\be\label{Vinv}
\vec{N}_{\D}^{\rm f}= V^{-1}\,\vec{N}_{\D''}^{\rm f} \, ,
\hspace{4mm} \mbox{\rm with}
\hspace{5mm}
V^{-1} \equiv \left(\begin{array}{ccc}
V^{-1}_{e e''} & V^{-1}_{\m e''} & V^{-1}_{\t e''} \\
V^{-1}_{e \m''} & V^{-1}_{\m \m''} & V^{-1}_{\t \m''} \\
V^{-1}_{e \t''} & V^{-1}_{\m \t''} & V^{-1}_{\t\t''}
\end{array}\right) \, ,
\ee
or explicitly for the single components
\bea \nonumber
N^{\rm f}_{\D_{\a}} & = & \sum_{\a''}\,V^{-1}_{\a\a''}\,
\left[N^{T\sim T_L}_{\a''}\,e^{-{3\pi\over 8}\,K_{1\a''}}\right] \\ \label{NfDa}
& = & \,
\sum_{\a''}\,V^{-1}_{\a\a''}\,\,e^{-{3\pi\over 8}\,K_{1\a''}}
\left[\sum_{\b}\,V_{\a''\b}\,N_{\D_{\b}}^{T\sim T_L}\right] \, ,
\eea
where the $N_{\D_{\b}}^{T\sim T_L}$'s are given by eqs.~(\ref{Dg}), (\ref{Dtau}) and (\ref{Demu}).
This equation is the general analytical solution and should be regarded
as the ``master equation'' of the paper.
It can be immediately checked
that taking $U=V=I$ one recovers the standard solution given by eq.~(\ref{finalasnoc}).
In the Appendix we recast it in an extensive way for illustrative purposes.

\section{Examples for strong impact of flavour coupling}

The general solution of eq.~(\ref{NfDa}), with approximate analytical solutions for $U$ and $V$ plugged in,
is of course rather lengthy and its physical implications are difficult to see.
To make  eq.~(\ref{NfDa}) more easily accessible we partly unpack it in the Appendix.
In order to better understand whether it can yield results
significantly different from those obtained by eq.~(\ref{finalasnoc}),
we will now specialize it to some interesting specific example
cases that will highlight the possibility of strong deviations
from the case when flavour coupling is neglected, i.e., of $R_{\rm f}$ (cf. (\ref{r})) values significantly
different from unity. The scenario we will consider in the following, and which will be useful to illustrate the possibility of large impact
of flavour coupling effects, will be referred to as the ``flavour-swap scenario''. Notice that in general
the phantom terms have to be taken into account and we have therefore included them. However, these can be always thought to vanish in the case
of initial vanishing abundance.

\subsection{Simplified formulae in the ``Flavour-swap scenario''}

In the ``flavour-swap scenario'' the following situation is considered: Out of the two flavours $e$ and $\m$, one has $K_{1\d}\lesssim 1$ (where $\d$ can be either $e$ or $\m$). The other flavour will be denoted by $\b$, so if $\d = e$ then $\b=\m$ or vice versa. For $K_{1\b}$ we will assume that $K_{1\b} \sim K_{1\t} \sim K_1 \gg 1$, such that asymmetries in the $\b''$ as well as in the $\t''$ flavours will be (almost) completely erased by the exponential $N_1$ washout. The only asymmetry relevant after $N_1$ washout will be the one in the flavour $\d''$.

Obviously, this already simplifies eq.~(\ref{NfDa}) significantly.
Now one has, similarly to what happened before with the $K_{1\a'}$,
that $K_{1\d''}= K_{1\d}\,(1+{\cal O}(C^{(3)}_{\a\neq\b})^3) \simeq K_{1\d}$. At the same time
$K_{1\b(\t)''}= K_{1\b(\t)}\,(1+{\cal O}(C^{(3)}_{\a\neq\b}))$ and therefore
$K_{1 \b(\t)''} \sim K_1 \gg 1$. This implies that in eq.~(\ref{NfDa}))
only the terms with $\a''=\d''$ survive
, while the terms with $\a''=\b'',\t''$
undergo a strong wash-out from the lightest RH neutrino inverse processes and can be
neglected. Therefore, if we calculate the final flavoured asymmetries
and make the approximation $\exp(-3\pi\,K_{1\d}/8)\simeq 1$, from the general
eq.~(\ref{appendix}) we can write
\bea
N^{\rm f}_{\D_\b} & \simeq & V^{-1}_{\b\d''}\,V_{\d'' \b}\,N_{\D_{\b}}^{T\sim T_L}+
                 V^{-1}_{\b \d''}\,V_{\d'' \d}\,N_{\D_{\d}}^{T\sim T_L}+
                   V^{-1}_{\b \d''}\,V_{\d'' \t}\,N_{\D_{\t}}^{T\sim T_L} \, , \\
N^{\rm f}_{\D_\d} & \simeq & V^{-1}_{\d \d''}\,V_{\d'' \b}\,N_{\D_{\b}}^{T\sim T_L}+
                 V^{-1}_{\d \d''}\,V_{\d'' \d}\,N_{\D_{\d}}^{T\sim T_L}+
                   V^{-1}_{\d \d''}\,V_{\d'' \t}\,N_{\D_{\t}}^{T\sim T_L} \, , \\
N^{\rm f}_{\D_\t} & \simeq & V^{-1}_{\t \d''}\,V_{\d'' \b}\,N_{\D_{\b}}^{T\sim T_L}+
                 V^{-1}_{\tau \d''}\,V_{\d'' \d}\,N_{\D_{\d}}^{T\sim T_L}+
                   V^{-1}_{\tau \d''}\,V_{\d'' \t}\,N_{\D_{\t}}^{T\sim T_L} \, ,
\eea
At the production, for the three $N_{\D_{\a}}^{T\sim T_L}$'s,
we assume the conditions that led to the
eqs.~(\ref{flasspec}), (\ref{flasspec2})  and (\ref{Demu}),
i.e. $P_{2\t}^0 \ll P_{2\g}^0$  (notice again that
one could also analogously consider the opposite case $P_{2\t}^0 \ll P_{2\g}^0$) and $K_2\gg 1$,
implying $\k(K_{2\g})\ll 1$.
The matrices $V$ and $V^{-1}$, whose entries are defined by the eqs.~(\ref{V}) and (\ref{Vinv}) respectively,
at the first order in the $C^{(3)}$ off-diagonal terms,
are given by
\be
V \simeq \left(\begin{array}{ccc}
1 &  C^{(3)}_{e\m} & -C^{(3)}_{e\t}\,{P^0_{1 e}\over P^0_{1\t}}  \\
-C^{(3)}_{\m e} {P^0_{1\m}\over P^0_{1 e}} & 1 & -C^{(3)}_{\m \t} {P^0_{1\m}\over P^0_{1 \t}} \\
C^{(3)}_{\t e}  & C^{(3)}_{\t \m}  & 1
\end{array}\right)
\, ,
\,\,\, V^{-1} \simeq \left(\begin{array}{ccc}
1 &  -C^{(3)}_{e\m} & C^{(3)}_{e\t}\,{P^0_{1 e}\over P^0_{1\t}} \\
 C^{(3)}_{\m e} {P^0_{1\m}\over P^0_{1 e}} & 1 &  C^{(3)}_{\m \t} {P^0_{1\m}\over P^0_{1 \t}} \\
-C^{(3)}_{\t e}  & -C^{(3)}_{\t \m} & 1
\end{array}\right)
\, .
\ee

Therefore, we find for the three $N_{\D_{\a}}^{\rm f}$'s
\bea
N^{\rm f}_{\D_\b} & \simeq & -C^{(3)}_{\b\d}\,C^{(3)}_{\d \b}\,{P^0_{1\d}\over P^0_{1 \b}}\,N_{\D_{\b}}^{T\sim T_L}
                 -C^{(3)}_{\b\d}\,N_{\D_{\d}}^{T\sim T_L}+
                  C^{(3)}_{\b\d}\,C^{(3)}_{\d \t} {P^0_{1\d}\over P^0_{1 \t}}\,N_{\D_{\t}}^{T\sim T_L} \\ \nonumber
                 & \simeq & -C^{(3)}_{\b\d}\,
                 \left\{p_\d+{P^0_{2\d}\over P^0_{2\g}}\,\left[\ve_{2\g}\k(K_{2\g})-\, C^{(2)}_{\g\t}\,\ve_{2\t} \,\k(K_{2\t})\right]\right\} \, , \\
N^{\rm f}_{\D_\d} & \simeq & -C^{(3)}_{\d \b} {P^0_{1\d}\over P^0_{1 \b}}\,N_{\D_{\b}}^{T\sim T_L}+
                 N_{\D_{\d}}^{T\sim T_L}-
                  C^{(3)}_{\d \t}\, {P^0_{1\d}\over P^0_{1 \t}} \,N_{\D_{\t}}^{T\sim T_L} \\ \nonumber
                 & \simeq &
                 p_\d+{P^0_{2\d}\over P^0_{2\g}}\,\left[\ve_{2\g}\k(K_{2\g})-\, C^{(2)}_{\g\t}\,\ve_{2\t} \,\k(K_{2\t})\right]
                  - C^{(3)}_{\d\t}\,{P^0_{1\d}\over P^0_{1 \t}} \,\ve_{2\t}\,\k(K_{2\t}) \, , \\
N^{\rm f}_{\D_\t} & \simeq & C^{(3)}_{\t \d}\, C^{(3)}_{\d \b} {P^0_{1\d}\over P^0_{1 \b}}\,N_{\D_{\b}}^{T\sim T_L}-
                C^{(3)}_{\t \d}\,N_{\D_{\d}}^{T\sim T_L}-
                   C^{(3)}_{\t \d}\,C^{(3)}_{\d \t} {P^0_{1\d}\over P^0_{1 \t}} \,N_{\D_{\t}}^{T\sim T_L} \\ \nonumber
                 & \simeq &
                 -\,C^{(3)}_{\t\d}\,\left\{p_\d+{P^0_{2\d}\over P^0_{2\g}}\,\left[\ve_{2\g}\k(K_{2\g})-\, C^{(2)}_{\g\t}\,\ve_{2\t} \,\k(K_{2\t})\right]\right\} \, .
\eea
The total final asymmetry is then given by the sum of the flavoured asymmetries.
It  can be checked
that if flavour coupling is neglected ($C^{(2)}=C^{(3)}=I$), then one obtains the expected result
\be
N_{B-L}^{\rm f}\simeq N_{\D_{\d}}^{T\sim T_L}=p_\d+{P^0_{2\d}\over P^0_{2\g}}\,\ve_{2\g}\k(K_{2\g}) \, ,
\ee
corresponding to an asymmetry produced in the flavour $\d$, i.e.\ in the only flavour that survives washout by the lightest RH neutrino.

However, taking into account flavour coupling, new terms arise and the
final asymmetry can be considerably enhanced. More explicitly, we have approximately
\be\label{eq:NBLflavourswap}
N_{B-L}^{\rm f}\simeq \left(1-C^{(3)}_{\b\d}-C^{(3)}_{\t\d}\right)
        \left\{p_\d+{P^0_{2\d}\over P^0_{2\g}}\,\left[\ve_{2\g}\k(K_{2\g})-\, C^{(2)}_{\g\t}\,\ve_{2\t} \,\k(K_{2\t})\right]\right\}
                 - C^{(3)}_{\d\t}\,{P^0_{1\d}\over P^0_{1 \t}}\,\ve_{2\t}\,\k(K_{2\t}) \, ,
\ee
where ${P^0_{1\d}/ P^0_{1 \t}} ={K_{1\d}/ K_{1 \t}} $ and where we have neglected all terms that contain the product either of two or more off-diagonal terms of the coupling matrix, or of one or more off-diagonal term with $\k(K_{2\g})\ll 1$.

From eq.~(\ref{eq:NBLflavourswap}) one can readily see examples for strong enhancement of the
asymmetries due to flavour coupling, i.e. conditions under which $R^{\rm f}\gg 1$.  In particular,
if $\k(K_{2\g})\,\ve_{2\g} \ll \k(K_{2\t})\,\ve_{2\t}$
then one of the two additional terms in eq.~(\ref{eq:NBLflavourswap}),
only present due to flavour coupling, can dominate the produced final asymmetry and $R^{\rm f}\gg 1$ results.
We will now discuss these two cases in more detail and give examples for classes of models, consistent with the observed neutrino masses and mixings, where they are relevant. We want first to notice a few general
things.

First, since the flavoured asymmetries are upper bounded by \cite{flavoreffects2}
\be\label{eps2aub}
|\ve_{2\a}|\lesssim \ve_{2\a}^{\rm max}\equiv
10^{-6}\,{M_2\over 10^{10}\,{\rm GeV}}\,\sqrt{P_{2\a}}\,{m_3\over m_{\rm atm}} \, ,
\ee
the condition $\k(K_{2\g})\,\ve_{2\g} \ll \k(K_{2\t})\,\ve_{2\t}$ does not introduce
further great restrictions compared to $K_{2\tau}\ll K_{2\gamma}$.
Second, from the eq.~(\ref{eq:NBLflavourswap}) one can see that a reduction
of the final asymmetry from flavour coupling is also possible because of a possible
sign cancelation among the different terms (in addition to a small reduction
from the pre-factor $1-C^{(3)}_{\b\d}-C^{(3)}_{\t\d}$). However, a strong reduction occurs
only for a fine tuned choice of the parameters. Let us say that this sign cancelation
introduced by flavour coupling changes the condition for the vanishing of the final asymmetry
that is not anymore simply given by $\ve_{2\gamma}=0$.

It should indeed be noticed that now for $\ve_{2\gamma}=0$
the asymmetry in the flavour $\gamma$ (or vice-versa the asymmetry in the
flavour $\tau$ if $\ve_{2\tau}=0$ and $K_{2\tau}\gg K_{2\gamma}$) does not vanish in general.
This can be seen directly from the kinetic equations (cf. eq.(\ref{flke})),
where if $\ve_{2\gamma}=0$ an asymmetry generation can be still induced
by the wash-out term that actually in this case behaves rather like a wash-in
 term. If we we focus on the Higgs asymmetry, we can say that this wash-in effect
is induced by a sort of thermal contact between the flavour $\gamma$ and $\tau$, in a way that
the departure from equilibrium in the flavour $\tau$ induces a departure from equilibrium
in the flavour $\gamma$ as well.

\begin{itemize}

\item {\bf Case A: Enhancement from flavour coupling at $N_2$ decay}

Let us assume  $\k(K_{2\g})\ll \k(K_{2\t})$ and in addition ${P^0_{1\d}/ P^0_{1 \t}} ={K_{1\d}/ K_{1 \t}}\ll 1 $.
Then the first and third terms in eq.~(\ref{eq:NBLflavourswap}) dominate and we can estimate
\be\label{eq:NBLflavourswap_caseA}
N_{B-L}^{\rm f}\simeq
p_\d-C^{(2)}_{\g\t}\,{P^0_{2\d}\over P^0_{2\g}}\,\ve_{2\t} \,\k(K_{2\t}) \, .
\ee
In this case the final asymmetry is dominated by two terms that, for different reasons,
circumvent the strong wash-out of the $\g$ component. The first term in eq.~(\ref{eq:NBLflavourswap_caseA}) is the
phantom term $p_\d$ that escapes the wash-out since it was `hidden' within the
coherent $\g$ lepton combination of an electron and a muon component. From this point of view
it should be noticed that since the lightest RH neutrino wash-out acts only on the
$\d$ flavour but not on the $\beta$ flavour, it has the remarkable effect
to destroy the cancelation between the two phantom terms $p_\d$ and $p_\b$ having
as a net effect the creation of $B-L$ asymmetry, a completely new effect.
The second term in eq.~(\ref{eq:NBLflavourswap_caseA}) is what we have seen already:
because of flavour coupling at the production, the large asymmetry in the $\t$ flavour
necessarily induces an asymmetry
in the $\g$ flavour as well. Notice that there is no model independent reason why one of
the two terms should dominate over the other.

In order to show more clearly the conditions for this case to be realized, we have plotted
in the Fig.~\ref{fig:case A} the $R$ iso-contour lines (cf. eq~(\ref{r})) in the plane $(K_{2\gamma},K_{2\tau})$.
\begin{figure}
\begin{center}
\psfig{file=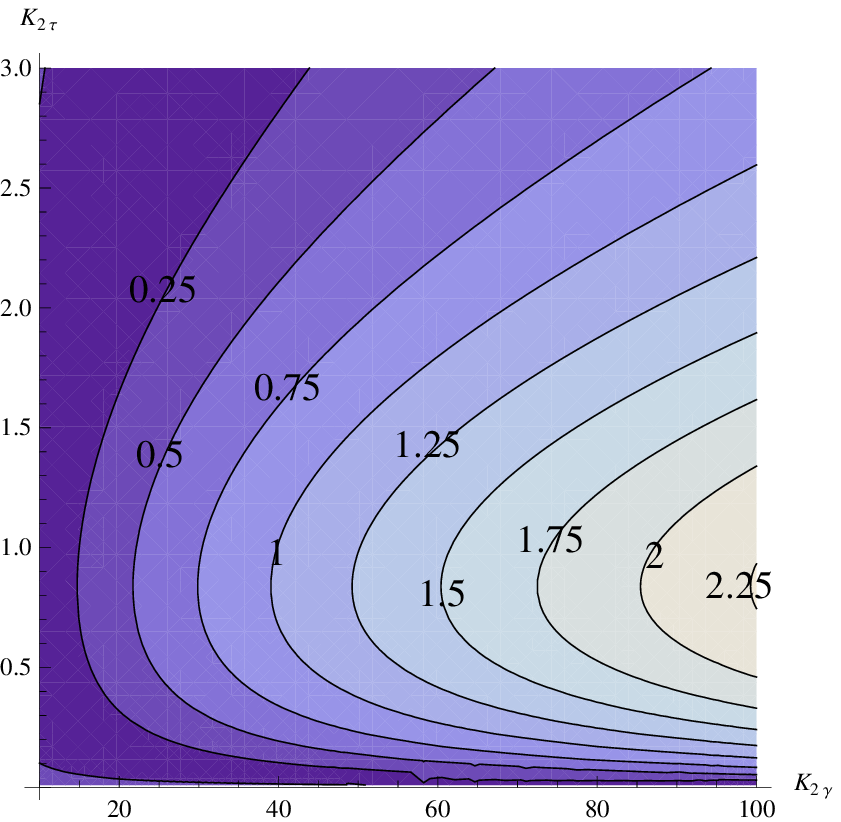,height=63mm,width=75mm}
\hspace{5mm}
\psfig{file=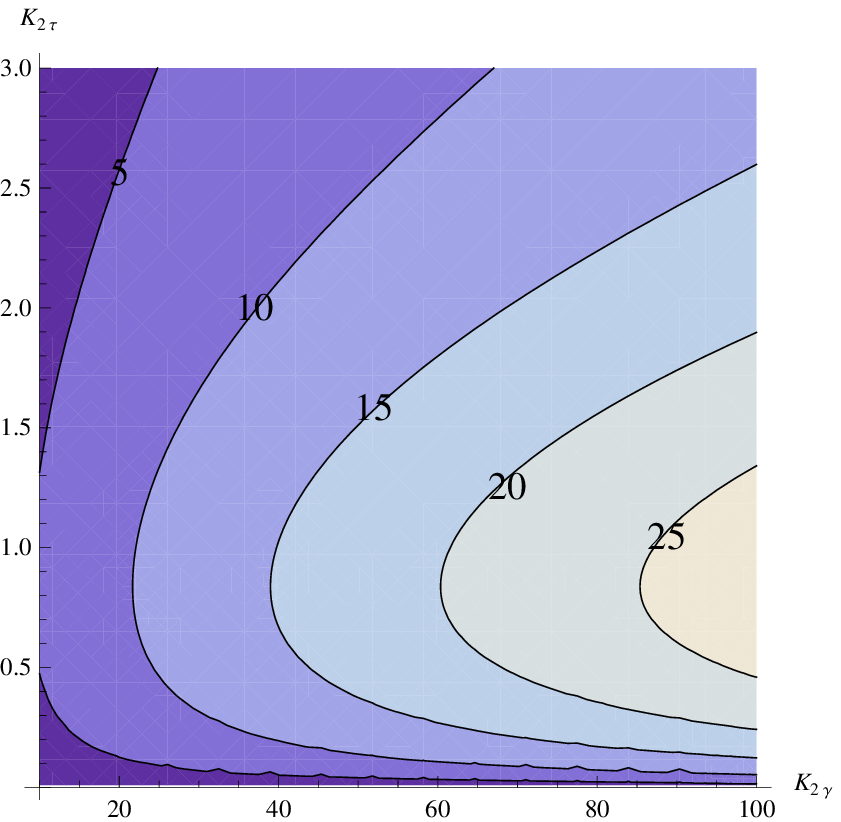,height=63mm,width=75mm}
\caption{Contour plots of $R$ (cf.\ eq~(\ref{r})) in the flavour swap scenario for
$K_{1\tau},K_{1 e}\gg 1$, $K_{1\mu}\lesssim 1$, $K_{2e}=K_{2\m}$.
The latter condition implies that the last term in the
eq.~(\ref{eq:NBLflavourswap}) is negligible. Left panel: $|\ve_{2\mu}|=\ve_{2\mu}^{\rm max}$;
right panel: $|\ve_{2\mu}|=0.1\,\ve_{2\mu}^{\rm max}$ (cf.\ eq.~(\ref{eps2aub})).
In both panels $\ve_{2\tau}=\ve_{2\tau}^{\rm max}$ and $\ve_{2\mu}/\ve_{2\tau}>1$.}
\label{fig:case A}
\end{center}
\end{figure}
 We have fixed
$K_{1\mu}\lesssim 1$, $K_{1e},K_{1\tau}\gg 1$, so that only the muonic asymmetry survives the
lightest RH neutrino wash-out. We have also set
$K_{2\mu}/K_{2\gamma}=1/2 \gg K_{1\mu}/K_{1\tau}$, so that the last term in the eq.~(\ref{eq:NBLflavourswap})
can be neglected. Concerning the $C\!P$ asymmetries, in the left panel
we have set $\ve_{2\gamma}=\ve_{2\g}^{\rm max}$ and $\ve_{2\t}=\ve_{2\t}^{\rm max}$.
One can see that in this case the enhancement of the asymmetry becomes relevant when $K_{2\gamma}\gg K_{2\tau}$
but for $K_{2\gamma}\lesssim 100$ (a reasonable maximum value), it cannot be higher than about $R\simeq 2.5$.
Notice that, since we choose $\ve_{2\gamma}/\ve_{2\t} > 1$, a reduction is also
possible due to a cancelation of the traditional term and of the new term due to flavour coupling.
In the right panels we have set $\ve_{2\gamma}=0.1\,\ve_{2\gamma}^{\rm max}$
and this time one can see how $R$ can be as large as one order of magnitude. This shows
that for $\ve_{\g}\rightarrow 0$ the enhancement can be arbitrarily large.

\item {\bf Case B: Enhancement from flavour coupling at $N_1$ washout}

Another interesting case is when $\k(K_{2\g})\ll \k(K_{2\t})$ and in addition
$P^0_{2\d}/P^0_{2\g} \ll P^0_{1\d}/P^0_{1\t} $.
In this case the first and fourth terms in  eq.~(\ref{eq:NBLflavourswap}) dominate and we obtain approximately
\be\label{eq:NBLflavourswap_caseB}
N_{B-L}^{\rm f}\simeq p_{\d}- C^{(3)}_{\d\t}\,{P^0_{1\d}\over P^0_{1 \t}}\,\ve_{2\t}\,\k(K_{2\t})
 \, .
\ee
\end{itemize}
We can see that again we have the phantom term avoiding the wash-out at the production
and a second term arising from the flavour coupling at the wash-out by $N_1$.
We note that this term is not even proportional to the
flavoured asymmetry $\ve_{2\d}$ and is just due to the fact that thanks to
flavour coupling the wash-out of the large tauonic asymmetry produced at $T\sim T_L$
has as a side effect a departure from thermal equilibrium of the processes
$N_1 \leftrightarrow l_e + \phi^{\dagger}, \bar{l}_e + \phi$. This can be understood
easily again in terms of the Higgs asymmetry that connects the dynamics in the two flavours.
It is quite amusing that thanks to flavour coupling an electron asymmetry is generated
even without explicit electronic $C\!P$ violation.

\begin{figure}
\begin{center}
\psfig{file=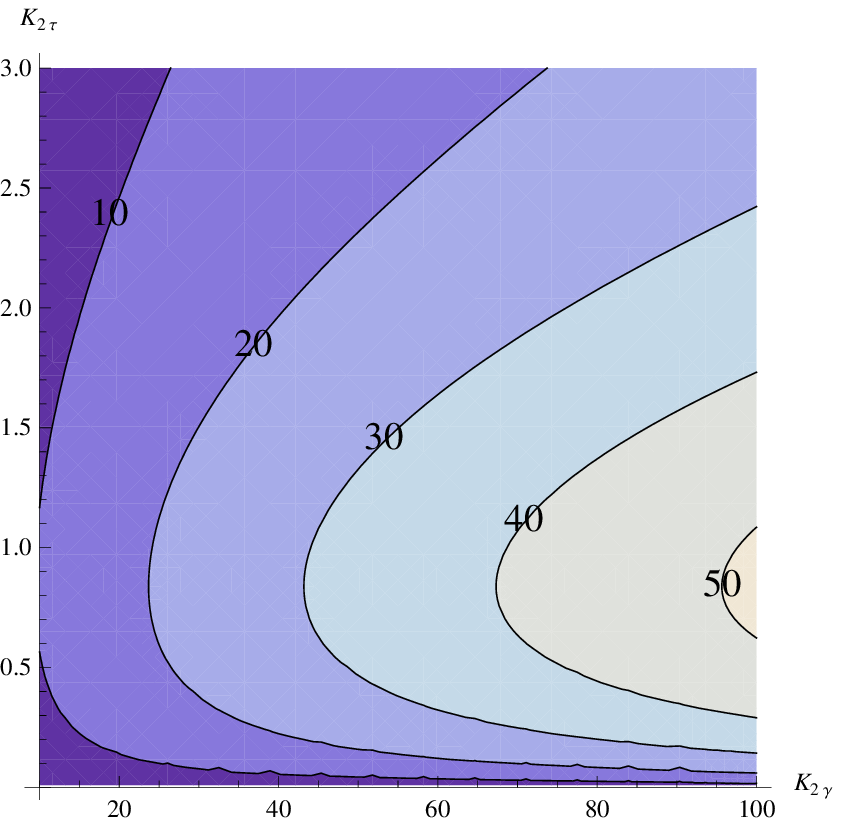,height=63mm,width=75mm}
\hspace{5mm}
\psfig{file=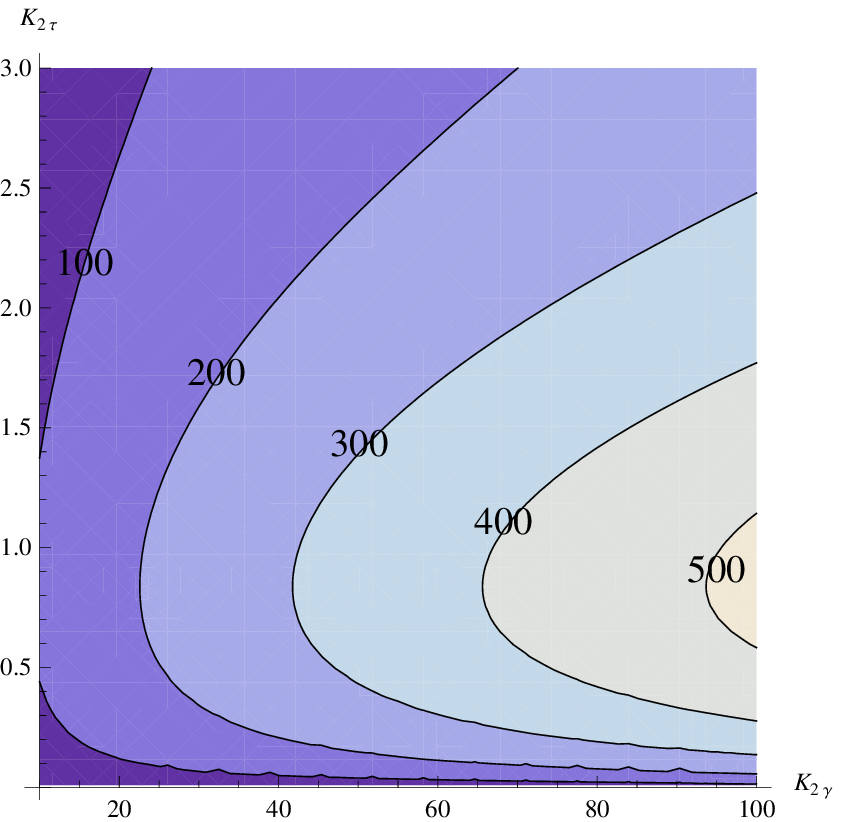,height=63mm,width=75mm}
\caption{Contour plots of $R$ (cf.\ eq~(\ref{r})) in the flavour swap scenario for
$K_{1\tau},K_{1 \mu}\gg 1$, $K_{1 e}\lesssim 1$, $K_{2e}/K_{2\m}\ll K_{1e}/K_{1\tau}$.
The last condition implies that the last term in the
eq.~(\ref{eq:NBLflavourswap}) dominates. Left panel: $\ve_{2\mu}=\ve_{2\mu}^{\rm max}$;
right panel: $\ve_{2\mu}=0.1\,\ve_{2\mu}^{\rm max}$ (cf.\ eq.~(\ref{eps2aub})).
In both panels $\ve_{2\tau}=\ve_{2\tau}^{\rm max}$ and $\ve_{2\mu}/\ve_{2\tau}>1$.}
\label{fig:case B}
\end{center}
\end{figure}

Also for this case B, we have plotted,
in the Fig.~\ref{fig:case B}, the $R$ iso-contour lines
(cf. eq~(\ref{r})) in the plane $(K_{2\gamma},K_{2\tau})$.
We have set $K_{1e}\lesssim 1$ while $K_{1\mu},K_{1\tau}\gg 1$, so that now only the electron
asymmetry survives the lightest RH neutrino wash-out.
Moreover this time we have set
$K_{2e}/K_{2\gamma} \ll  K_{1e}/K_{1\tau}$ so that the last term
in the eq.~(\ref{eq:NBLflavourswap}) becomes dominant and the case B is realized.
For the $C\!P$ asymmetries, as before, in the left panel
we fixed $\ve_{2\gamma}=\ve_{2\g}^{\rm max}$
while in the right panel $\ve_{2\gamma}=0.1\,\ve_{2\g}^{\rm max}$ and in both cases $\ve_{2\t}=\ve_{2\t}^{\rm max}$.
Now the enhancement of the final asymmetry $R$ is $\gg 1$ in both cases, simply because
the traditional term is this time suppressed by $K_{2e}/K_{2\gamma} \ll 1$. This means that
after the decoherence of the $\gamma$ lepton quantum states, there is a negligible asymmetry in the electron flavour.
However, at the lightest RH neutrino wash-out, an electron asymmetry is generated thanks to
flavour coupling.

\subsection{Example for Case A within Heavy Sequential Dominance}

To find realistic examples where the two cases A and B with strong impact of flavour coupling are realised, we will now consider classes of models with so-called sequential dominance (SD) \cite{King:1998jw,King:1999mb,King:2002nf,King:2004} in the seesaw mechanism.
To illustrate case A, we may in particular consider a sub-class called heavy sequential dominance (HSD). To realise case A within HSD, in eq.~(\ref{eq:NBLflavourswap}) and eq.~(\ref{eq:NBLflavourswap_caseA}) we assign flavours $\d = \m$ and $\b = e$.

To understand how heavy sequential dominance works, we begin by
writing the RH neutrino Majorana mass matrix $M_{\mathrm{RR}}$ in
a diagonal basis as
\begin{equation}
M_{\mathrm{RR}}=
\begin{pmatrix}
M_C & 0 & 0 \\
0 & M_B & 0 \\
0 & 0 & M_A%
\end{pmatrix},
\end{equation}
where we have ordered the columns
according to $M_{RR}=\mbox{diag}(M_1,M_2,M_3)$ where $M_1<M_2<M_3$.
In this basis we write the neutrino (Dirac) Yukawa matrix $\lambda_{\nu}$ in
terms of $(1,3)$ column vectors $C_i,$ $B_i,$ $A_i$ as
\begin{equation}
\lambda_{\nu }=
\begin{pmatrix}
C & B & A
\end{pmatrix},
  \label{Yukawa}
\end{equation}
in the convention where the Yukawa matrix is given in left-right convention.
The Dirac neutrino mass matrix is then given by $m_{\mathrm{LR}}^{\nu}=\lambda_{\nu}v_{\mathrm{
u}}$. The term for the light neutrino masses in the effective Lagrangian (after electroweak symmetry breaking), resulting from integrating out the massive right
handed neutrinos, is
\begin{equation}
\mathcal{L}^\nu_{eff} = \frac{(\nu_{i}^{T} A_{i})(A^{T}_{j} \nu_{j})v^2}{M_A}+\frac{(\nu_{i}^{T} B_{i})(B^{T}_{j} \nu_{j})v^2}{M_B}
+\frac{(\nu_{i}^{T} C_{i})(C^{T}_{j} \nu_{j})v^2}{M_C}  \label{leff}
\end{equation}
where $\nu _{i}$ ($i=1,2,3$) are the left-handed neutrino fields.
heavy sequential dominance (HSD) then corresponds to the third
term being negligible, the second term subdominant and the first term
dominant:
\beq\label{SDcond}
\frac{A_{i}A_{j}}{M_A} \gg
\frac{B_{i}B_{j}}{M_B} \gg
\frac{C_{i}C_{j}}{M_C} \, .
\eeq
In addition, we shall shortly see that small $\theta_{13}$
and almost maximal $\theta_{23}$ require that
\beq
|A_1|\ll |A_2|\approx |A_2|.
\label{SD2}
\eeq
We identify the dominant
RH neutrino and Yukawa couplings as $A$, the subdominant
ones as $B$, and the almost decoupled (subsubdominant) ones as $C$.

Working in the mass basis
of the charged leptons,
we obtain for the lepton mixing angles:
\begin{subequations}\label{anglesSD}\begin{eqnarray}
\label{Eq:t23}
\tan \theta_{23} &\approx& \frac{|A_2|}{|A_3|}\;, \\
\label{Eq:t12}
\tan \theta_{12} &\approx&
\frac{|B_1|}{c_{23}|B_2|\cos \tilde{\phi}_2 -
s_{23}|B_3|\sin \tilde{\phi}_3  } \;,\\
\label{Eq:t13}
\theta_{13} &\approx&
e^{i \tilde{\phi}_4}
\frac{|B_1| (A_2^*B_2 + A_3^*B_3) }{\left[|A_2|^2 + |A_3|^2\right]^{3/2} }
\frac{M_A}{M_B}
+\frac{e^{i \tilde{\phi}_5} |A_1|}
{\sqrt{|A_2|^2 + |A_3|^2}} ,
\end{eqnarray}\end{subequations}
where the phases do not need to concern us.

The neutrino masses are:
\begin{subequations}\label{massesSD}\begin{eqnarray}
\label{Eq:m3} m_3 &\approx& \frac{(|A_2|^2 + |A_3|^2)v^2}{M_A}\;, \\
\label{Eq:m2} m_2 &\approx& \frac{|B_1|^2 v^2}{s^2_{12} M_B}\;, \\
\label{Eq:m1}m_1 &\approx& {\cal O}(|C|^2 v^2/M_C) \;.
\end{eqnarray}\end{subequations}

Tri-bimaximal mixing corresponds to:
\begin{eqnarray}
|A_{1}| &=&0,  \label{tribicondsd} \\
\text{\ }|A_{2}| &=&|A_{3}|,  \label{tribicondse} \\
|B_{1}| &=&|B_{2}|=|B_{3}|,  \label{tribicondsa} \\
A^{\dagger }B &=&0.  \label{zero}
\end{eqnarray}
This is called constrained sequential dominance (CSD).

For $N_2$ leptogenesis, the flavour specific decay asymmetries are $\varepsilon_{2 \alpha}$ where the leading contribution comes from the heavier RH neutrino of mass $M_A=M_3$ in the loop which may be approximated via eq.~(\ref{eps2a}) as:
\be
\varepsilon_{2 \alpha} \approx -\frac{3 }{16 \pi v^2} \frac{M_2}{M_3}\frac{1}{B^\dagger B}
\mathrm{Im}\left[ B_\alpha^* (B^\dagger A)A_\alpha  \right].
\ee
Clearly the asymmetry vanishes in the case of CSD due to eq.~(\ref{zero})
and so in the following we shall consider examples which violate CSD.
The mixing angles are given by the following estimates:
\be\label{eq:angles}
 \tan \theta_{23}\sim \frac{A_2}{A_3} \sim 1, \ \ \tan \theta_{12}\sim \frac{\sqrt{2}B_1}{B_2+B_3}\sim \frac{1}{\sqrt{2}},
 \ \ \theta_{13}\sim \frac{A_1}{\sqrt{2}A_2} \sim \frac{r}{\sqrt{2}}.
\ee
Suppose we parametrize the Yukawa couplings consistent with these mixing angles as:
\be
A_2=A_3, \ \ A_1=r\,A_2, \ \  B_3=q\,B_2,\ \  B_1=\frac{1}{2} (1+q)\,B_2\ \
\ee
where $r<1$ is related to $\theta_{13}$ and $\theta_{12}$ via eq.~(\ref{eq:angles}), then we find,
\be
\varepsilon_{2 \mu} \approx -\frac{3 }{16 \pi v^2}\,M_2m_3, \ \ \varepsilon_{2 \tau}\approx q\, \varepsilon_{2 \mu},
\ \ \varepsilon_{2 e}\approx \frac{r}{2}\,\varepsilon_{2\mu}.
\ee
The flavoured effective neutrino masses $\widetilde{m}_{2 \alpha}$, $\widetilde{m}_{1 \alpha}$ are given by:
\be
\widetilde{m}_{2 \alpha} = \frac{|B_{\alpha}|^2 v^2}{M_B} \sim m_2 \, , \, \,
\widetilde{m}_{1 \alpha}= \frac{|C_{\alpha}|^2 v^2}{M_C}\sim m_1 \, .
\ee
Neutrino oscillation experiments tell us that $r<1$ is small (here we shall assume $r\sim 0.2$  as
a specific example consistent with current experimental results) and we find
\be
K_{2 \mu }={\widetilde{m}_{2 \mu} \over m_{\star}} \sim {m_2\over m_{\star}} \sim 10, \, \,
K_{2 e} \sim \frac{(1+q)^2}{4}\,K_{2\mu }, \, \, K_{2 \tau}\sim q^2\,K_{2 \mu } ,
\ee
which allows strong washout for $K_{2 \gamma}$ ($\gamma = \mu + e$) with weak washout for $K_{2 \tau}$.
By assuming that $C_{1},C_2 \ll C_3$ we have,
\be
K_{1 \tau}= {\widetilde{m}_{1 \tau} \over m_{\star}} \sim 10\,\frac{m_1}{m_2}, \, \, K_{1 e}, \, K_{1 \mu} \ll  K_{1 \tau}
\ee
which allows for strong washout for $K_{1 \tau}$
(at least if $m_1\sim m_2$) with weak washouts for $K_{1 e}, K_{1 \mu}$.

Thus, without flavour coupling and phantom terms,
we would have strong (exponential) $N_1$ washout for $K_{1 \tau}\sim 10$,
with negligible $N_1$ washout for $K_{1e}, K_{1 \mu}<1$.
Since $\varepsilon_{2 e}\approx \frac{r}{2} \varepsilon_{2 \mu} \sim 0.1 \varepsilon_{2 \mu}$ we may
neglect $\varepsilon_{2 e}$ and then we find
that the term proportional to $\ve_{2 \g}\,\kappa (K_{2 \gamma})$ is strongly
washed out since $K_{2 \gamma}\sim 10$. Therefore, without flavour coupling and
phantom effects, $N_{B-L}^{\rm f}$ would tend to be small in this scenario.

\begin{figure}
\begin{center}
\psfig{file=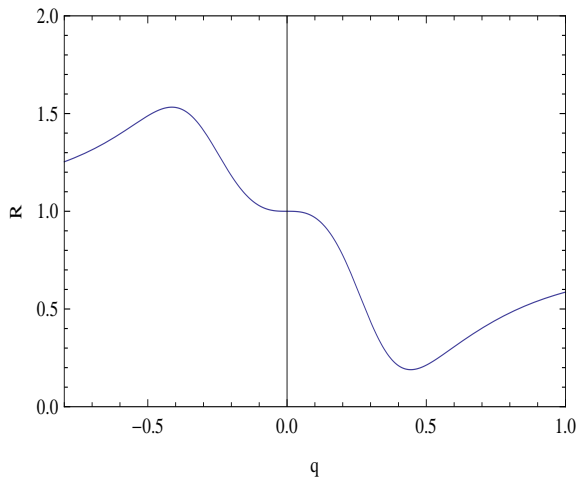,height=63mm,width=75mm}
\caption{Plot of $R$ as a function of $q$ as from the eq.~(\ref{Rq}).}
\label{fig:R}
\end{center}
\end{figure}

While, allowing for the effects of flavour redistribution and including the phantom
term, we find (cf. eq.~(\ref{eq:NBLflavourswap_caseA})),
\be
N_{B-L}^{\rm f}\sim p_\mu + {K_{2 \m}\over K_{2 \g}}\,\varepsilon_{2 \gamma}\,\kappa(K_{2\gamma})
- {K_{2 \m}\over K_{2 \g}}
C^{(2)}_{\gamma \tau}\,\varepsilon_{2 \tau}\,\kappa(K_{2 \tau}) \, .
\ee
Since $K_{2 \m}/K_{2 \g} \simeq 4/(5+2\,q)$ and
$p_{\m}\simeq [(1+2q)/(5+2q)]\ve_{2 \m}\,N_{N_2}^{\rm in}$, then we have
\be
N_{B-L}^{\rm f}\sim {1+2q\over 5+2q}\,\ve_{2 \m}\,N_{N_2}^{\rm in}+ {4\over 4+(1+q)^2}\,
\left[ \varepsilon_{2 \gamma}\,\kappa(K_{2\gamma})  -
\,C^{(2)}_{\gamma \tau} \varepsilon_{2 \t}\, \kappa(K_{2 \tau})\right] \, ,
\ee
where $K_{2 \tau}\sim q^2\,K_{2 \mu }\sim 10\,q^2 $ leads to only weak wash out with
$\varepsilon_{2 \mu}\sim -\frac{3 }{16 \pi v^2}M_2\,m_3$ being large. Notice that
there is a partial cancelation of  the two terms
but this is just depending  on the particular choice of values for $r$ and $q$ and on $N_{N_2}^{\rm in}$.
This is an example, consistent with neutrino data, where $N_{B-L}^{\rm f}$ would be very small without flavour coupling and phantom term, but will be quite large including the two effects that both
produce a large contribution. If we indeed, for definiteness, assume $N_{N_2}^{\rm in}=0$
and $q\sim 0.5$ such that $K_{2\tau}\sim 1$ corresponding to $\kappa(K_{2\tau})\simeq 0.3$,
then we find for $R$ (cf. eq~(\ref{r}))
\be\label{Rq}
R \simeq \left|1-C^{(2)}_{\gamma \tau}\,{\kappa(K_{2 \tau})\over \kappa(K_{2\gamma})}\,
{\varepsilon_{2\,\t} \over \varepsilon_{2 \gamma}}\right|  \, .
\ee
In Fig.~(\ref{fig:R}) we plotted $R$ as a function of $q=\ve_{2\tau}/\ve_{2\mu}$. One can see
that this example realizes a specific case of the general situation shown in the left panel of
Fig.~1. In particular, one can see that there can be a relevant suppression for
positive $q$ and up to a $50\%$ enhancement for negative $q$.
On the other hand, in case of initial thermal abundance, one can easily
verify that the presence of the phantom term can yield an enhancement up to three orders of magnitude.

\subsection{Example for Case B within Light Sequential Dominance}

To give an example for case B (i.e.\ an example where  $K_{1 e} \ll K_{1 \mu}, K_{1 \tau}$ while $\varepsilon_{2 \tau} \gg \varepsilon_{2 \mu}, \varepsilon_{2 e}$ and $K_{2 e}\ll K_{2 \gamma}$), we may consider another class of sequential dominance, namely light sequential dominance (LSD). Now, in eq.~(\ref{eq:NBLflavourswap}) and eq.~(\ref{eq:NBLflavourswap_caseB}) we have to replace $\d = e$ and $\b = \m$.

In the example of LSD we will consider, using the same notation for the dominant, subdominant and subsubdominant RH neutrinos and corresponding couplings, we have:
\begin{equation}
M_{\mathrm{RR}}=
\begin{pmatrix}
M_A & 0 & 0 \\
0 & M_C & 0 \\
0 & 0 & M_B%
\end{pmatrix}.
\end{equation}
The lightest RH neutrino with mass $M_A$ dominates the seesaw mechanism.
We have again ordered the columns according to $M_{RR}=\mbox{diag}(M_1,M_2,M_3)$ where $M_1<M_2<M_3$.
For the neutrino (Dirac) Yukawa matrix we use the notation
\begin{equation}
\lambda_{\nu }=
\begin{pmatrix}
A_1 & C_1 & B_1 \\
A_2 & C_2 & B_2 \\
A_3 & C_3 & B_3
\end{pmatrix},.
  \label{YukawaLSD}
\end{equation}

More specifically, let us now consider, within LSD, a variant of CSD called partially constrained sequential dominance (PCSD) \cite{King:2009qh} where $|A_2| = |A_3| = a$ and $|B_1|=|B_2|=|B_3|=b$, but $A_1 \not= 0$. In addition, we may assume $C = (C_1,C_2,C_3)$ with $C_1 = 0$ and $C_2 / C_3 = \zeta \ll 1$ as a specific example. Under these conditions, and using $A_1 = r A_2 = \sqrt{2} \theta_{13} A_2$ defined in the previous section,
we can write the neutrino Yukawa matrix as
\begin{equation}
\lambda_{\nu }=
\begin{pmatrix}
\sqrt{2} \theta_{13} a  & 0 & b \\
a  & \zeta c & b\\
-a & c & b
\end{pmatrix}.
  \label{YukawaPCSDinLSD}
\end{equation}

The flavoured effective neutrino masses
$\widetilde m_{2\alpha}$, $\widetilde m_{1\alpha}$ in this specific LSD scenario are given by:
\be
\widetilde m_{2\alpha} = \frac{|C_{\alpha}|^2 v^2}{M_C} \sim m_2, \ \
\widetilde m_{1\alpha}= \frac{|A_{\alpha}|^2 v^2}{M_A}\;.
\ee
For $\widetilde m_{1 e}$, $\widetilde m_{1 \mu} $ and $\widetilde m_{1 \tau}$ we obtain explicitly
\be
\widetilde m_{1 e}= \frac{|\sqrt{2}\, \theta_{13}\, a|^2 v^2}{M_1} = m_3\, \theta_{13}^2, \ \
\widetilde m_{1 \mu} = \widetilde m_{1 \tau} =  \frac{|a|^2 v^2}{M_1} = \frac{m_3}{2} \;.
\ee
 The parameters $K_{i\alpha}$ are related to the $\widetilde m_{i \alpha}$'s
 simply by $K_{i\alpha}=\widetilde m_{i \alpha}/m^*$.
Since we know from neutrino oscillation experiments that the leptonic mixing angle $\theta_{13}$ is small
(at least $< 10^\circ$) we have that $K_{1 e} \ll K_{1 \mu} = K_{1 \tau}$, i.e.\
\begin{equation}
K_{1 \mu} = K_{1 \tau} \sim {m_3\over  m^*} \sim 50
\end{equation}
and
\begin{equation}
\frac{K_{1 e}}{K_{1 \mu}} = \frac{K_{1 e}}{K_{1 \tau}} =   (\sqrt{2} \theta_{13})^2 \;.
\end{equation}
Consequently, the asymmetries in the $\tau$ and in the $\mu$ flavours will be almost completely washed out by the $N_1$ washout related to $K_{1 \tau}$ and $K_{1 \mu}$. In the $e$-flavour we have weak $N_1$-washout.

Furthermore, using $\frac{|C_{\alpha}|^2 v^2}{M_C} \sim m_1$, we obtain at the $N_2$ decay stage
\begin{equation}
K_{2 \tau} \sim {m_1 \over  m^*} , \ \  K_{2 \mu} \sim \zeta\, {m_1 \over  m^*} \ll  K_{2 \tau}, \   \mbox{and} \, \,
K_{2 e} = 0 \:,
\end{equation}
which implies
\begin{equation}
K_{2 \g} = K_{2 \mu}+ K_{2 e} \ll  K_{2 \tau} \:.
\end{equation}

The $N_2$ decay asymmetries,  ignoring the contribution with $N_1$ in the loop which is very small for the considered case that $N_1 \ll N_2$, are given via eq.~(\ref{eps2a}) by
\be
\varepsilon_{2 \alpha} \approx -\frac{3 }{16 \pi v^2} \frac{M_2}{M_3}\frac{1}{B^\dagger B}
\mathrm{Im}\left[ B_\alpha^* (B^\dagger C)C_\alpha  \right].
\ee
Using $B$ and $C$ as specified above eq.~(\ref{YukawaPCSDinLSD}) and $m_1 \sim \frac{|C_{\alpha}|^2 v^2}{M_C}$, we obtain for the decay asymmetries $\varepsilon_{2 \a}$:
\be
\varepsilon_{2\tau} \sim -\frac{3 }{16 \pi v^2}\,M_2\, m_2, \ \
\varepsilon_{2\mu} = \zeta \varepsilon_{2\tau} \ll \varepsilon_{2\tau}, \ \
\varepsilon_{2e} = 0\;.
\ee

Considering eq.~(\ref{eq:NBLflavourswap}) and noting that $K_{2 e} = 0$ together with $\varepsilon_{2 e} = 0$ implies $p_\delta = 0$ we see that all terms apart from the one proportional to $C^{(3)}_{e\t}$ are strongly suppressed provided that $\zeta$ is sufficiently tiny ($\zeta \ll r$). In other words, the considered LSD scenario provides an example for case B, a final asymmetry
dominated by flavour coupling effects at the $N_1$ washout stage, as in eq.~(\ref{eq:NBLflavourswap_caseB}). Explicitly, we obtain for the final asymmetry
\be
N_{B-L}^{\rm f} \sim
 - C^{(3)}_{e\t}\,{K_{1 e}\over K_{1 \t}}\,\ve_{2\t}\,\k(K_{2 \t})
\sim
 \frac{3 C^{(3)}_{e\tau}}{16 \pi } \frac{M_2\, m_2}{v^2 } (\sqrt{2} \theta_{13})^2 \k\left(\frac{m_1}{m^*} \right) .
\ee
Here one can see that
\be
R \simeq 1+0.01\,\zeta^{-1}\,\left({\theta_{13}\over 10^\circ}\right)^2 \,
{\kappa(m_1/m_{\star})\over 0.3} \, .
\ee
This result shows quite interestingly that, if $\theta_{13}\neq 0$ and $m_1\gtrsim m_{\star}$,
one can obtain a huge enhancement for $\xi\rightarrow 0$, indicating that,
accounting for flavour coupling, one can have an asymmetry in a situation where
one would otherwise obtain basically a zero asymmetry. This happens because part of the
tauon asymmetry, thanks to  flavour coupling at the lightest RH neutrino wash-out,
escapes the wash out from the lightest RH neutrinos.

\section{Conclusions}
We have discussed various new flavour dependent effects in the $N_2$-dominated scenario of leptogenesis
and have shown that these effects are important in obtaining a reliable expression for the final asymmetry.
 In particular we have emphasized the importance of the off-diagonal entries of the flavour coupling matrix that
  connects the total flavour asymmetries, distributed in different particle species,
  to the lepton and Higgs doublet asymmetries. We have derived analytical formulae for the final asymmetry
  including the flavour coupling at the $N_2$-decay stage, where effectively two flavours are active,
  as well as at the stage of washout by the lightest
  RH neutrino $N_1$ where all three flavours are distinguished.
  The interplay between the production
stage and the wash-out stage can then result in a significant enhancement of the
final asymmetry.

We have also described a completely new effect,
``phantom leptogenesis'', where the lightest RH neutrino wash-out is actually able to create
a $B-L$ asymmetry rather than destroying it as usually believed. This is possible because
the individual wash-out on each flavoured asymmetry can erase cancelations among the
electron and muon phantom terms and therefore lead to a net increase of the total $B-L$ asymmetry.
In this way the wash-out at the production is basically fully circumvented for part of the
produced electron and muon asymmetries. We also noticed however that the phantom
terms also strongly depend on the specific initial conditions since they are proportional
to the initial $N_2$-abundance and therefore, in particular, they vanish for initial zero $N_2$-abundance.

The changes induced by these new effects are encoded in the general ``master formula''
eq.~(\ref{NfDa}) for the final asymmetry that we derived from the Boltzmann equations
without approximations.
Based on this equation we have identified a sufficiently generic scenario, the ``flavour swap scenario'',
where we proved that a strong enhancement of the final asymmetry due to flavour
coupling and phantom terms is clearly possible. The conditions for the flavour swap scenario
correspond to have a one flavour dominated asymmetry at the production, in the two flavour regime,
and a wash-out from the lightest RH neutrinos swapping the dominance  from one flavour to the other.
Flavour coupling  can strongly modify the flavour asymmetry that is subdominant at the production
inducing two new contributions, one generated at the production and one at the lightest RH neutrino
wash-out. Then, in the flavour swap scenario, this translates into a strong modification of the final asymmetry
after the lightest RH neutrino wash-out.
It is quite interesting that, because of flavour coupling, an asymmetry is
actually generated by the wash-out terms that therefore in this case act more
like wash-in terms, transmitting a departure from thermal equilibrium from
one flavour to the other. In the figures we have showed how, once the flavour swap scenario,
is realized, relevant modifications of the final asymmetry are generically induced by flavour coupling.
Depending on the values of the involved parameters,
these range from ${\cal O}(1)$ factor changes (either a reduction or an enhancement) to
an orders of magnitude enhancement.

We have illustrated these effects for two models which describe
realistic neutrino masses and mixing  based on sequential dominance.

In conclusion, the off-diagonal flavour
couplings as well as phantom terms can have a significant impact on the baryon asymmetry produced
by $N_2$-dominated leptogenesis and thus have to be included in a reliable analysis. We have derived exact
analytic (and also compact approximate) results that allow this to be achieved. The results in this paper open up
new possibilities for successful $N_2$-dominated leptogenesis to explain the baryon asymmetry of the universe.

\subsection*{Acknowledgments}
S.A.\  acknowledges partial support from the DFG cluster of excellence ``Origin and Structure
of the Universe''. P.D.B. acknowledges financial support from the NExT Institute and SEPnet.
S.F.K.\ was partially supported by the following grants:
STFC Rolling Grant ST/G000557/1 and a Royal Society Leverhulme Trust Senior Research
Fellowship. P.D.B wishes to thank Antonio Riotto and Steve Blanchet for useful discussions.

\section*{Appendix A}
\appendix

\renewcommand{\thesection}{\Alph{section}}
\renewcommand{\thesubsection}{\Alph{section}.\arabic{subsection}}
\def\theequation{\Alph{section}.\arabic{equation}}
\renewcommand{\thetable}{\arabic{table}}
\renewcommand{\thefigure}{\arabic{figure}}
\setcounter{section}{1}
\setcounter{equation}{0}

In this Appendix we recast eq.~(\ref{NfDa}) in a more extensive way
in order to illustrate a generic feature of it. Each final $\alpha$ asymmetry
is now the sum of three contributions,
\bea\label{NfDa2}
N^{\rm f}_{\D_{\a}}
 & = &  V^{-1}_{\a e''}\,
\left[\sum_{\b}\,V_{e''\b}\,N_{\D_{\b}}^{T\sim T_L}\right]
\,e^{-{3\pi\over 8}\,K_{1 e''}} \\  \nonumber
& + &  V^{-1}_{\a \m''}\,
\left[\sum_{\b}\,V_{\m''\b}\,N_{\D_{\b}}^{T\sim T_L}\right]
\,e^{-{3\pi\over 8}\,K_{1 \m''}} \\ \nonumber
& + &  V^{-1}_{\a \t''}\,
\left[\sum_{\b}\,V_{\t''\b}\,N_{\D_{\b}}^{T\sim T_L}\right]
\,e^{-{3\pi\over 8}\,K_{1 \t''}} \, .
\eea
In the approximation $K_{1 \e''}\simeq K_{1 e}$,
$K_{1 \m''}\simeq K_{1 \m}$, $K_{1 \t''}\simeq K_{1 \t}$,
it becomes
\bea\label{NfDa2}
N^{\rm f}_{\D_{\a}}
 & \simeq &  V^{-1}_{\a e''}\,
\left[\sum_{\b}\,V_{e''\b}\,N_{\D_{\b}}^{T\sim T_L}\right]
\,e^{-{3\pi\over 8}\,K_{1 e}} \\  \nonumber
& + &  V^{-1}_{\a \m''}\,
\left[\sum_{\b}\,V_{\m''\b}\,N_{\D_{\b}}^{T\sim T_L}\right]
\,e^{-{3\pi\over 8}\,K_{1 \m}} \\ \nonumber
& + &  V^{-1}_{\a \t''}\,
\left[\sum_{\b}\,V_{\t''\b}\,N_{\D_{\b}}^{T\sim T_L}\right]
\,e^{-{3\pi\over 8}\,K_{1 \t}} \, .
\eea
This expression shows how now each $\alpha$ asymmetry
is not simply given by one term containing a $N_1$ wash-out exponential
suppression term  described by $e^{-3\pi\,K_{1\a}/8}$ but it also contains terms
that are washed out by exponentials $e^{-3\pi\,K_{1\d\neq\a}/8}$. In this way,
even though $K_{1\a}\gg 1$, there can still be unsuppressed contributions
to $N^{\rm f}_{\D_{\a}}$ from terms with $K_{1\d\neq\a}\ll 1$. Even though
these terms are weighted by factors $V^{-1}_{\a\d}$
containing off-diagonal terms of the $C^{(3)}$ matrix, they
can be dominant in some cases and therefore, in general, they have to
be accounted for.

We can also recast this last equation in even a more explicit form
unpacking the second sum as well,
\bea\label{appendix}
N^{\rm f}_{\D_{\a}}
 & \simeq &  V^{-1}_{\a e''}\,
\left[V_{e'' e}\,N_{\D_{e}}^{T\sim T_L}+V_{e''\m}\,N_{\D_{\m}}^{T\sim T_L}+
V_{e''\t}\,N_{\D_{\t}}^{T\sim T_L} \right]
\,e^{-{3\pi\over 8}\,K_{1 e}} \\  \nonumber
& + &  V^{-1}_{\a \m''}\,
\left[V_{\m'' e}\,N_{\D_{e}}^{T\sim T_L}+V_{\m''\m}\,N_{\D_{\m}}^{T\sim T_L}+
V_{\m''\t}\,N_{\D_{\t}}^{T\sim T_L} \right]
\,e^{-{3\pi\over 8}\,K_{1 \m}} \\ \nonumber
& + &  V^{-1}_{\a \t''}\,
\left[V_{\t'' e}\,N_{\D_{e}}^{T\sim T_L}+V_{\t''\m}\,N_{\D_{\m}}^{T\sim T_L}+
V_{\t''\t}\,N_{\D_{\t}}^{T\sim T_L} \right]
\,e^{-{3\pi\over 8}\,K_{1 \t}} \, .
\eea

\section*{Appendix B}
\appendix

\renewcommand{\thesection}{\Alph{section}}
\renewcommand{\thesubsection}{\Alph{section}.\arabic{subsection}}
\def\theequation{\Alph{section}.\arabic{equation}}
\renewcommand{\thetable}{\arabic{table}}
\renewcommand{\thefigure}{\arabic{figure}}
\setcounter{section}{2}
\setcounter{equation}{0}
In this Appendix we show how the results on phantom leptogenesis can be also derived within a density matrix formalism.
We report results from \cite{preparation}, where a more general and detailed discussion can be found.
Let us recall that we treat the three stages of $N_2$ production at $T\sim T_L$,  decoherence  at $T^{\star}\sim 10^9\,$GeV
and the lightest RH neutrino wash-out at $T\sim M_1\ll 10^9\,$GeV, as completely separate from each other. This allows a
simplification of the discussion. In order to simplify the notation, in this Appendix we also assume that
the quantum states $|\ell_2\rangle$ and $|\bar{\ell}'_2\rangle$ do not have any tauon component, i.e.
$P_{2\tau}=\bar{P}_{2\tau}=P^0_{2\tau}=0$ and consequently $P_{2\gamma}=\bar{P}_{2\gamma}=P^0_{2\gamma}=1$.

The flavour composition of the  lepton and anti-lepton quantum states  can then be written as $(\a=e,\mu)$
\bea
|{\ell}_2\rangle & = &
{\cal C}_{2e}\,|{\ell}_e \rangle + {\cal C}_{2\mu}\,|{\ell}_\mu \rangle \, , \;\;\;
{\cal C}_{2\a} \equiv  \langle {\ell}_\a|\ell_2 \rangle  \, , \\
|\bar{\ell}'_2\rangle & = &
\bar{{\cal C}}_{2e}\,|\bar{{\ell}}_e \rangle +  \bar{{\cal C}}_{2\mu}\,|\bar{{\ell}}_\mu\rangle \, , \;\;\;
\bar{{\cal C}}_{2\a} \equiv  \langle \bar{{\ell}}_\a|\bar{\ell}'_2 \rangle \, ,
\eea
where $P_{2\a}=|{\cal C}_{2\a}|^2$ and  $\bar{P}_{2\a}=|\bar{\cal C}_{2\a}|^2$.
At the $N_2$-production, at $T\sim T_L \gg 10^9\,{\rm GeV}$,
muon charged lepton interactions  are ineffective and therefore the
$|\ell_2\rangle$ and $|\bar{\ell}'_2\rangle$ quantum states evolve coherently. In this case,
in the two different bases  $\ell_2-\ell_2^\bot$ and $\bar{\ell}'_2-\bar{\ell}_2^{'\bot}$, the
lepton and anti-lepton density matrices are respectively simply given by 
\be
\rho^{\ell_2}_{ij} \equiv   \langle i|{\ell}_2\rangle\langle {\ell}_2| j\rangle  ={\rm diag}(1,0)  \hspace{5mm} \mbox{\rm and}
 \hspace{5mm}
\rho^{\bar{\ell}'_2}_{\bar{i}\bar{j}}  \equiv  \langle \bar{i}|\bar{\ell}'_2\rangle\langle \bar{\ell}'_2|\bar{j}\rangle ={\rm diag}(1,0) \, ,
\ee
where $|i\rangle ,|j\rangle = |{\ell_2}\rangle, |{\ell^{\bot}_2}\rangle$ and 
$|\bar{i}\rangle ,|\bar{j}\rangle = |{\bar{\ell}'_2}\rangle, |{\bar{\ell}^{'\bot}_2}\rangle$.
It is crucial to notice that,
because of the different flavour composition of $|\ell_2\rangle$ and $|\bar{\ell}'_2\rangle$,
the two bases are not $C\!P$ conjugated of each other.
If we introduce the lepton number and anti-lepton number density matrices,
$N^{\ell_2}_{ij}\equiv N_{\ell_2} \, \rho^{\ell_2}_{ij}$ and
$N^{\bar{\ell}'_2}_{\bar{i}\bar{j}}\equiv N_{\ell_2} \, \rho^{\bar{\ell}'_2}_{\bar{i}\bar{j}}$
respectively, their evolution at $T\sim T_L$ is given by
 \be\label{dmke}
{dN^{\ell_2}_{ij}\over dz_2}  =
\left({\G_2\over H\,z_2}\,N_{N_2}- {\G_2^{ID}\over H\,z_2}\,N_{\ell_2} \right)\,\rho^{\ell_2}_{ij}  \, , \;\; \hspace{5mm}
{dN^{\bar{\ell}'_2}_{\bar{i}\bar{j}}\over dz_2}  =
\left({\bar{\G}_2\over H\,z_2}\,N_{N_2}- {\bar{\G}_2^{ID}\over H\,z_2}\,N_{\bar{\ell}_2} \right)
\,\rho^{\bar{\ell}'_2}_{\bar{i}\bar{j}} \, .
\ee
In order to obtain an equation for the total $B-L$ asymmetry $N_{B-L}$ at $T\simeq T_L$,
we have first to write these two equations in the same flavour basis for leptons and anti-leptons,
that for our objectives can be conveniently chosen  to be $e-\mu$ and $\bar{e}-\bar{\mu}$ respectively, 
and then subtract them. Introducing the rotation operators $R$ and $\hat{R}$,
such that $|{\ell}_2\rangle  = \hat{R}\,|e\rangle$, $|{\ell}^{\bot}_2\rangle  = \hat{R}\,|\mu\rangle$ 
and $|\bar{\ell}'_2\rangle  = \hat{\bar{R}}\,|\bar{e}\rangle$,  $|\bar{\ell}^{'\bot}_2\rangle  = \hat{\bar{R}}\,|\bar{\mu}\rangle$,
the corresponding rotations matrices $R_{i\alpha}$ and $\bar{R}_{i\a}$ are 
\be
R_{i\a} = \left(\begin{array}{cc}
{\cal C}_{2e} & -{\cal C}_{2\m} \\
{\cal C}_{2\m} &  {\cal C}_{2e}
\end{array}\right)  \hspace{10mm}\mbox{\rm and} \hspace{10mm}
\bar{R}_{\bar{i}\bar{\a}} = \left(\begin{array}{cc}
\bar{\cal C}_{2e} & -\bar{\cal C}_{2\m} \\
\bar{\cal C}_{2\m} &  \bar{\cal C}_{2e}
\end{array}\right)
\ee
for leptons and anti-leptons respectively. In the lepton flavour basis
one can finally write the equation for the $B-L$ asymmetry density matrix
\be\label{dNBmLab}
{dN^{B-L}_{\a\b} \over dz_2} =\bar{R}^{\dagger}_{\bar{\a}\bar{i}}\,{dN^{\bar{\ell}_2}_{\bar{i}\bar{j}}\over dz_2}\,\bar{R}_{\bar{j}\bar{\b}} -
R^{\dagger}_{\a i}\,{dN^{\ell_2}_{ij}\over dz_2} \, R_{j\b}  \, .
\ee
For the relevant diagonal components one finds
\bea\label{dNBmLee}
{dN^{B-L}_{ee}\over dz_2} & = &
\ve_{2e}\,D_2\,(N_{N_2}-N_{N_2}^{\rm eq})-
P_{2e}^{0}\,W_2\,N_{B-L}  \, , \\ \label{dNBmLmm}
{dN^{B-L}_{\m\m}\over dz_2} & = &
\ve_{2\m}\,D_2\,(N_{N_2}-N_{N_2}^{\rm eq})-
P_{2\m}^{0}\,W_2\,N_{B-L}   \, .
\eea
Summing these two equations, one finds the usual
equation for the total $B-L$ asymmetry $N_{B-L}={\rm Tr}[N^{B-L}_{\a\b}]$,
\be
{dN_{B-L}\over dz_2}  =
\ve_{2}\,D_2\,(N_{N_2}-N_{N_2}^{\rm eq})- W_2\,N_{B-L}  \, ,
\ee
that is washed-out at the production. On the other hand, from
the two eqs.~(\ref{dNBmLee}) and (\ref{dNBmLmm}), one also obtains
\be
{1\over P^0_{2e}}\,{dN^{B-L}_{ee}\over dz_2}-
{1\over P^0_{2\m}}\,{dN^{B-L}_{\m\m}\over dz_2}
=-{\Delta P^0_{2e}\over 2}\,\left({1\over P^0_{2e}}+{1\over P^0_{2\m}}\right)\,D_2\,(N_{N_2}-N_{N_2}^{\rm eq}) \,
\ee
and from this finally
\be
N^{B-L}_{ee}(T\simeq T_L) = P^0_{2e}\,N_{B-L}^{T\simeq T_L} - {\D P_{2e}\over 2} \, N_{N_2}^{\rm in} \,  , \,\,
N^{B-L}_{\m\m}(T\simeq T_L) \simeq  P^0_{2\m}\,N_{B-L}^{T\simeq T_L} +   {\D P_{2e}\over 2}\, N_{N_2}^{\rm in} \, .
\ee
If we assume that $N_{B-L}$ is so strongly washed-out that the
terms  $\propto N_{B-L}^{T\simeq T_L}$ can be neglected, one has
$N^{B-L}_{ee}(T\simeq T_L) \simeq - (\D P_{2e}/ 2) \, N_{N_2}^{\rm in} \simeq -   N^{B-L}_{\m\m}(T\simeq T_L)$.
At this stage the electron and muon asymmetries are not measured and cannot have any 
physical consequence. 

However afterwards, at $T \sim T^{\star}=10^9\,$GeV, the coherence of the lepton quantum states
is broken by the muon lepton interactions and the electron and muon asymmetries
are measured by the thermal bath. One can show that
the charged lepton interactions give rise to additional terms in the eq.~(\ref{dNBmLab}) \cite{decoherence2}
that have the effect to damp the off-diagonal terms in $N^{B-L}_{\a\b}$.  The electron
and muon asymmetries are reprocessed by sphaleron processes and they correspond to 
$\D_e$ and $\D_\m$ asymmetries, $N_{\D_e}^{T_L}=N^{B-L}_{ee}$,  $N_{\D_\m}^{T_L}=N^{B-L}_{\m\m}$
that are now measured by the  thermal bath.  Nevertheless, at $T\sim T^{\star}$
and down to $T \gtrsim M_1$, they still do not give any contribution to the baryon asymmetry
(in this sense they are phantom) since they still cancel with each other.
However, after the $N_1$-wash out at $T\lesssim M_1$, one finally has
\be
N_{B-L}^{\rm f} \simeq - {\D P_{2e}\over 2} \, N_{N_2}^{\rm in}\,e^{-{3\pi\over 8}\,K_{1e}}
                                      +{\D P_{2e}\over 2} \, N_{N_2}^{\rm in}\,e^{-{3\pi\over 8}\,K_{1\m}}
\ee
that is nothing else than the eq.~(\ref{finalasnoc}) specialized to the case
$P^0_{2\g}=\ve_{2\tau}=\k(K_{2\g})=0$ (it is just a simple exercise to
include a tauon component re-obtaining exactly the eq.~(\ref{finalasnoc})).
Finally if, for example, one assumes $K_{1e}\lesssim 1$ and $K_{1\m}\gg 1$ ,
then $N_{B-L}^{\rm f} \simeq N_{\D_e}^{\rm f} \simeq - \D P_{2e} \, N_{N_2}^{\rm in}$:
the unwashed phantom term gets finally imprinted into the final asymmetry.
 It is worth to notice again that phantom leptogenesis has the attractive
feature to provide a way to circumvent completely the wash-out at the production but, on the other hand, it
also has the drawback that the final asymmetry strongly depends on the initial $N_2$ abundance.
Therefore, we have shown how phantom leptogenesis can be also described within a density matrix formalism
reproducing the results obtained in the main text in the Boltzmann formalism.
The actual practical advantage of the density matrix formalism is
that it allows to extend the description beyond the limit where the three stages
are fully separated, in a way that the final asymmetry can be calculated for 
a generic choice of the RH neutrino masses \cite{preparation}.
Notice that these results extend those obtained in \cite{decoherence2,beneke}  within
$N_1$ leptogenesis, to the case where more  heavy neutrino flavours contribute
to the final asymmetry. In  \cite{decoherence2,beneke} the lepton 
and anti-lepton density matrices were assumed to be diagonalizable  in  bases 
that are $C\!P$ conjugated of each other and in this case phantom terms
cannot be derived.  However, within $N_1$ leptogenesis, this does not affect the final asymmetry 
since phantom terms cancel anyway.

\end{document}